\documentclass[prd,aps,epsfig,amsmath,amsfonts,showpacs,nofootinbib]{revtex4}

\usepackage{times}
\usepackage{hyperref}
\usepackage{pifont}
\usepackage{graphicx}
\usepackage{dcolumn}
\usepackage{bm}
\usepackage{slashed}
\usepackage{color} 

\begin{document}

\title{
Dirac form factors and electric charge radii of baryons in the combined chiral and $1/N_c$ expansions}

\author{
Rub\'en Flores-Mendieta
}

\affiliation{
Instituto de F{\'\i}sica, Universidad Aut\'onoma de San Luis Potos{\'\i}, \'Alvaro Obreg\'on 64, Zona Centro, San Luis Potos{\'\i}, S.L.P.\ 78000, Mexico
}

\author{
Mayra Alejandra Rivera-Ruiz
}
\affiliation{
Instituto de F{\'\i}sica, Universidad Aut\'onoma de San Luis Potos{\'\i}, \'Alvaro Obreg\'on 64, Zona Centro, San Luis Potos{\'\i}, S.L.P.\ 78000, Mexico
}

\date{\today}

\begin{abstract}
The baryon Dirac form factor is computed at one-loop order in large-$N_c$ baryon chiral perturbation theory, where $N_c$ is the number of color charges. Loop graphs with octet and decuplet intermediate states are systematically incorporated into the analysis and the effects of the decuplet-octet mass difference are accounted for. There are large-$N_c$ cancellations between different one-loop graphs as a consequence of the large-$N_c$ spin-flavor symmetry of QCD baryons. As a byproduct, the mean-square charge radius is also computed through a detailed numerical analysis. The predictions of large-$N_c$ baryon chiral perturbation theory are in very good agreement both with the expectations from the $1/N_c$ expansion and with the experimental data.
\end{abstract}


\maketitle

\section{Introduction}

Almost two decades ago, a $1/N_c$ expansion of the chiral Lagrangian for baryons was proposed \cite{jen96} to analyze the low-energy dynamics of baryons interacting with pseudoscalar mesons in a combined expansion in chiral symmetry breaking and $1/N_c$, where $N_c$ is the number of color charges. Since then, many static properties of baryons have been successfully studied, namely, flavor-$\mathbf{27}$ baryon mass splittings \cite{jen96}, axial-vector couplings \cite{rfm06,rfm12}, magnetic moments \cite{rfm09,rfm14}, and more recently, vector couplings \cite{rfm14b}. On general grounds, the theoretical expressions obtained in those analyses agree well with data, providing strong evidence that the combined chiral and $1/N_c$ expansions work well for baryons.

A comprehensive study of the static properties of baryons has led to a better understanding of their fundamental structure. Accounting for the finite size of baryons, which is a consequence of the confinement of quarks inside a finite spatial volume, still represents a challenge for hadron physics research. The electromagnetic interaction is indeed a unique tool to probe the size of baryons (and mesons, of course) via scattering processes. The analysis of the electromagnetic scattering of particles with internal structure introduces the concept of electromagnetic form factors, which describe the spatial distribution of the charge and magnetization density in the baryons. 

In particular, the study of the neutron intrinsic charge radius is a uniquely interesting task. Although the neutron is electrically neutral, it is composed of charged quarks so it may have an internal charge distribution. In that case, the neutron will undergo electrostatic interactions with other charged particles such as electrons. Experimentally, electron-neutron scattering has been so far the only reliable method of determining the value and sign of the intrinsic radius of the electric-charge distribution in the neutron. An extraction of this static property from electron-deuteron scattering is not very reliable because of difficulties implicit to the deuteron structure. Indeed, the authors of Ref.~\cite{part} suggest the value $\langle r_n^2\rangle=-0.1161 \pm 0.0022\,\,\mathrm{fm}^2$. The negative sign may be explained in different approaches. The simplest explanation is provided by the SU(6) quark model which predicts a vanishing electric form factor and, hence, a vanishing mean-square charge radius \cite{carlitz77}. SU(6) symmetry is broken by the spin-dependent quark interaction, which is responsible for the $\Delta$-$N$ mass splitting; the interaction would then pull the up-quark to the center of the neutron and push the down-quarks to the edge, which would result in a nonzero electric charge form factor and, consequently, into a negative mean-square charge radius \cite{isgur80}. In another approach based on the quark model, it is assumed that there is a repulsion between quarks with parallel spins, so the quark orbits are slightly distorted. This results into a slightly negative surface and a slightly positive center because the down-quarks are a bit farther from the center than the up-quark so the charges will not cancel at each distance from the center \cite{smith}.

Over the years, the electromagnetic structure of baryons has been studied analytically within several frameworks. An important selection of such frameworks can be found in the quark model and its variants (nonrelativistic, relativistic, chiral) \cite{krivo,sch,sch93b,buch1,wagner,dahiya}, the $1/N_c$ expansion \cite{jen94,jen02,buch00,buch02,buch03}, and chiral perturbation theory \cite{but93,ban95,bern98,kubis2001,puglia,pas04,pas07,arn03,hac05,tib09}. It is difficult to assess the success of the many calculations of corrections to the electromagnetic form factors. Predictions that vary substantially from one another are obtained. Lattice QCD, on the other hand, has made great inroads into the subject by evaluating from first principles the QCD contribution to the baryon charge \cite{noz90,lein92,cloet03,lee05,ale06,ale08,aub08,boi09,wang,ale10,shan14a,shan14b}. Lattice simulations directly at the physical pion mass are becoming available, so eventually chiral extrapolations will not be needed and one source of systematic uncertainty will be eliminated.

In this paper, the formalism of the $1/N_c$ expansion combined with heavy baryon chiral perturbation theory will be used to compute the one-loop corrections to the Dirac form factor. As a byproduct, the baryon charge radii will also be studied. The analysis builds on earlier works, particularly Refs.~\cite{rfm14} and \cite{rfm14b}, where the magnetic moment and baryon axial and vector couplings, respectively, have been obtained. However, since the same approach used in these references will be followed here, much of the work has already been advanced. Therefore, the SU(3) breaking corrections to the Dirac form factor will be given at the leading order of the breaking, which is order $\mathcal{O}(p^2)$. This is not a drawback, though, because the analysis will be useful to envisage a more refined calculation to order $\mathcal{O}(p^3)$, as it has already been dealt with, for instance, in Refs.~\cite{kubis2001,arn03,geng09}, in the context of some variants of chiral perturbation theory (relativistic, quenched, and covariant).

The paper is organized as follows. A brief outline of baryon electromagnetic form factors is presented in Sec.~\ref{sec:emff} in order to introduce elementary definitions and conventions. In Sec.~\ref{sec:oneloop}, after a brief description of the methodology, the one-loop corrections to the Dirac form factor are computed. There are four Feynman diagrams that contribute to order $\mathcal{O}(p^2)$, and they are evaluated individually. In Sec.~\ref{sec:totalc}, all four one-loop contributions along with tree-level contributions are put together in order to construct the most general expression for the Dirac form factor for both octet and decuplet baryons. The computation of the mean-square charge radius is straightforwardly performed afterwards and the full expressions are given in Sec.~\ref{sec:bcr} for the sake of completeness. In Sec.~\ref{sec:num}, some numerical evaluations are performed for consistency. First, a fit to the experimental data is performed in order to find the values of the free parameters; then, the different contributions that make up electric charge radii are displayed and plotted. The paper is closed with some remarks in Sec.~\ref{sec:rem}.

\section{\label{sec:emff}A survey of baryon electromagnetic form factors}

Some properties of the electromagnetic current $J_\mu^{\mathrm{em}}$ can be better understood by exploiting the fact that it is a symmetry current conserved by all known interactions. The hadronic charge operator
\begin{equation}
Q_h^{\mathrm{em}} = \int J_0^{\mathrm{em}} d^3x,
\end{equation}
obeys the Gell-Mann--Nishijima relation
\begin{equation}
Q_h^{\mathrm{em}} = I_z + \frac{Y}{2},
\end{equation}
where $I_z$ is the third component of isospin and $Y$ is the hypercharge. But $I_z$ and $Y$ are given in terms of the SU(3) generators $T^3$ and $T^8$ by
\begin{equation}
I_z = T^3, \qquad Y = \frac{2}{\sqrt{3}}T^8.
\end{equation}
Thus, the electric charge transforms as a member of an $\mathbf{8}$ representation of SU(3). It is sensible to assume that $J_\mu^{\mathrm{em}}$ has the same transformation properties as $Q_h^{\mathrm{em}}$, so it is the sum of two parts: one that transforms as an isotriplet and one that transforms as an isospin singlet, namely,
\begin{equation}
J^{\mathrm{em}}_{h\mu} = J^3_\mu + \frac12 J_\mu^Y.
\end{equation}

In the heavy baryon formalism \cite{jm255,jm259}, the matrix elements of the electromagnetic current between baryon octet states can be written as
\begin{equation}
\langle \overline{B}(p_2)|J_\mu^{\mathrm{em}}|B(p_1)\rangle = \overline{u}(p_2) \left[v_\mu F_1(q^2) + \frac{[S_\mu,S_\nu]}{M_B}q^\nu F_2(q^2) \right] u(p_1),
\end{equation}
where $u(p_i)$ is the spinor of the baryon with momentum $p_i$, $q^2=(p_2-p_1)^2>0$ is the momentum transfer squared, $S_\mu$ is the spin operator, and $F_1(q^2)$ and $F_2(q^2)$ are usually referred to as Dirac and Pauli form factors, respectively, normalized in such a way that $F_1(0)=Q_B$ and $F_2(0) =\kappa_B$, where $Q_B$, $\kappa_B$, and $M_B$ denote the baryon charge, anomalous magnetic moment, and mass, respectively.

From the experimental bent, it is more convenient to express the Dirac and Pauli form factors in terms of the electric and magnetic Sachs form factors $G_{E0}(t)$ and $G_{M1}(t)$, defined as
\begin{subequations}
\begin{eqnarray}
G_{E0}(t) & = & F_1(t) + \frac{t}{4M_B^2} F_2(t), \label{eq:sachs1} \\
G_{M1}(t) & = & F_1(t) + F_2(t), \label{eq:sachs2}
\end{eqnarray}
\end{subequations}
which describe the distributions of the electric charge and the magnetic current in the Breit frame, respectively. In electron scattering, $t = -q^2$ is negative.

For baryon decuplet states, the matrix elements are \cite{arn03,geng09}
\begin{equation}
\langle \overline{T}(p_2)|J_\mu^{\mathrm{em}}|T(p_1)\rangle = -\overline{u}^\alpha(p_2) O_{\alpha\mu\beta}u^\beta(p_1),
\end{equation}
where $u_\alpha(p_i)$ is a Rarita-Schwinger spinor for an on-shell heavy baryon and the tensor $O_{\alpha\mu\beta}$ is written in terms of four form factors, namely,
\begin{eqnarray}
O_{\alpha\mu\beta} = g_{\alpha\beta} \left[v_\mu F_1(q^2) + \frac{[S_\mu,S_\rho]}{M_B}q^\rho F_2(q^2) \right] + \frac{q_\alpha q_\beta}{4M_B^2} \left[v_\mu G_1(q^2) + \frac{[S_\mu,S_\rho]}{M_B}q^\rho G_2(q^2) \right]. \label{eq:mtxd}
\end{eqnarray}
The Dirac- and Pauli-like form factors introduced in Eq.~(\ref{eq:mtxd}) can be combined to make up the electric charge $G_{E0}(t)$, magnetic dipole $G_{M1}(t)$, electric quadrupole $G_{E2}(t)$, and magnetic octupole $G_{M3}(t)$ form factors. The relation of interest here is
\begin{subequations}
\begin{equation}
G_{E0}(t) = F_1(t) + \frac{t}{4M_B^2} F_2(t) - \frac{t}{6M_B^2} G_{E2}(t), \label{eq:sachs1d}
\end{equation}
\end{subequations}
where, at $q^2=0$, the electric quadrupole moment $\mathbb{Q}$ is defined by
\begin{equation}
\mathbb{Q} \equiv \frac{1}{M_B^2} G_{E2}(0)= \frac{1}{M_B^2} \left[ Q_B - \frac12 G_1(0) \right].
\end{equation}

\section{\label{sec:oneloop}One-loop corrections to the Dirac form factor}

In the limit of exact flavor SU(3) symmetry, the hadronic weak vector and axial-vector currents belong to SU(3) octets, so the form factors of different baryon semileptonic decays (BSD) are related by SU(3) flavor symmetry and are given in terms of a few effective form factors and some well-known Clebsch-Gordan coefficients. The weak currents and the electromagnetic current are members of the same SU(3) octet, so all the vector form factors of BSD are related at $q^2=0$ to the electric charges and the anomalous magnetic moments of the nucleons.

At $q^2=0$, the baryon matrix elements for the vector current are given by the matrix elements of the associated charge or SU(3) generator. The flavor octet baryon charge, $V^c$, is spin-0 and a flavor octet, so it transforms as $(0,\mathbf{8})$ under SU(2)$\times$SU(3). The matrix elements of $V^c$ between SU(6) symmetric states yield the values of the leading vector form factor $f_1$ of BSD. The flavor index $c$ is simply set to $c=1\pm i2$ and $c=4\pm i5$ for strangeness-conserving and strangeness-changing weak decays, respectively.

On the other hand, SU(3) symmetry breaking effects in $V^c$ have been analyzed using the $1/N_c$ expansion \cite{rfm98} and the combined chiral and $1/N_c$ expansions \cite{rfm14b}. In the latter case, the computation was performed at one-loop level and the full result to order $\mathcal{O}(p^2)$ in the chiral expansion was achieved. Some variants of chiral perturbation theory (relativistic, covariant), with or without decuplet baryon degrees of freedom, have been used to approach the problem to the next order, $\mathcal{O}(p^3)$ \cite{krause,al93,vil06,lkm07,geng09}.

Now, in order to tackle the Dirac form factor in the combined chiral and $1/N_c$ expansion also to $\mathcal{O}(p^2)$, the method proposed in Ref.~\cite{rfm14b} will be followed. One might naively think that setting the flavor index $c$ of the one-loop correction $\delta V^c$ to $3+(1/\sqrt{3})8$ will work. However, the matter is not quite that simple. Although in the large-$N_c$ limit the baryon vector couplings have the same kinematic properties as the electromagnetic form factors and can be written in terms of the same operators, some subtleties will arise.

The operator giving rise to the Dirac form factor is thus a scalar that does not change either the electric charge or the strangeness; it connects states with the same $J$, $I_z$ and $Y$, and even it can connect states with different $I$, keeping the other quantum numbers unchanged. At $q^2=0$, it must reduce to the baryon electric charge.

Thus, at zero recoil, the Dirac form factor is
\begin{equation}
F_{1,\mathrm{Tree}}(0) = \langle T^Q \rangle = Q_B,
\end{equation}
where 
\begin{eqnarray}
T^Q \equiv T^3 + \frac{1}{\sqrt{3}} T^8 = \left( \begin{array}{rrr} \frac23 & 0 & 0 \\ 0 & -\frac13 & 0 \\ 0 & 0 & -\frac13
\end{array}
\right),
\end{eqnarray}
is the charge matrix for the three light quarks $u$, $d$, $s$. Hereafter, the flavor index $Q$ will stand for $Q=3+(1/\sqrt{3})8$ so any operator of the form $X^Q$ should be understood as $X^3+(1/\sqrt{3})X^8$. Similarly, $\overline{Q}$ will stand for $Q=3-(1/\sqrt{3})8$ so any operator of the form $X^{\overline{Q}}$ should be understood as $X^3-(1/\sqrt{3})X^8$.

\begin{figure}[t]
\scalebox{0.45}{\includegraphics{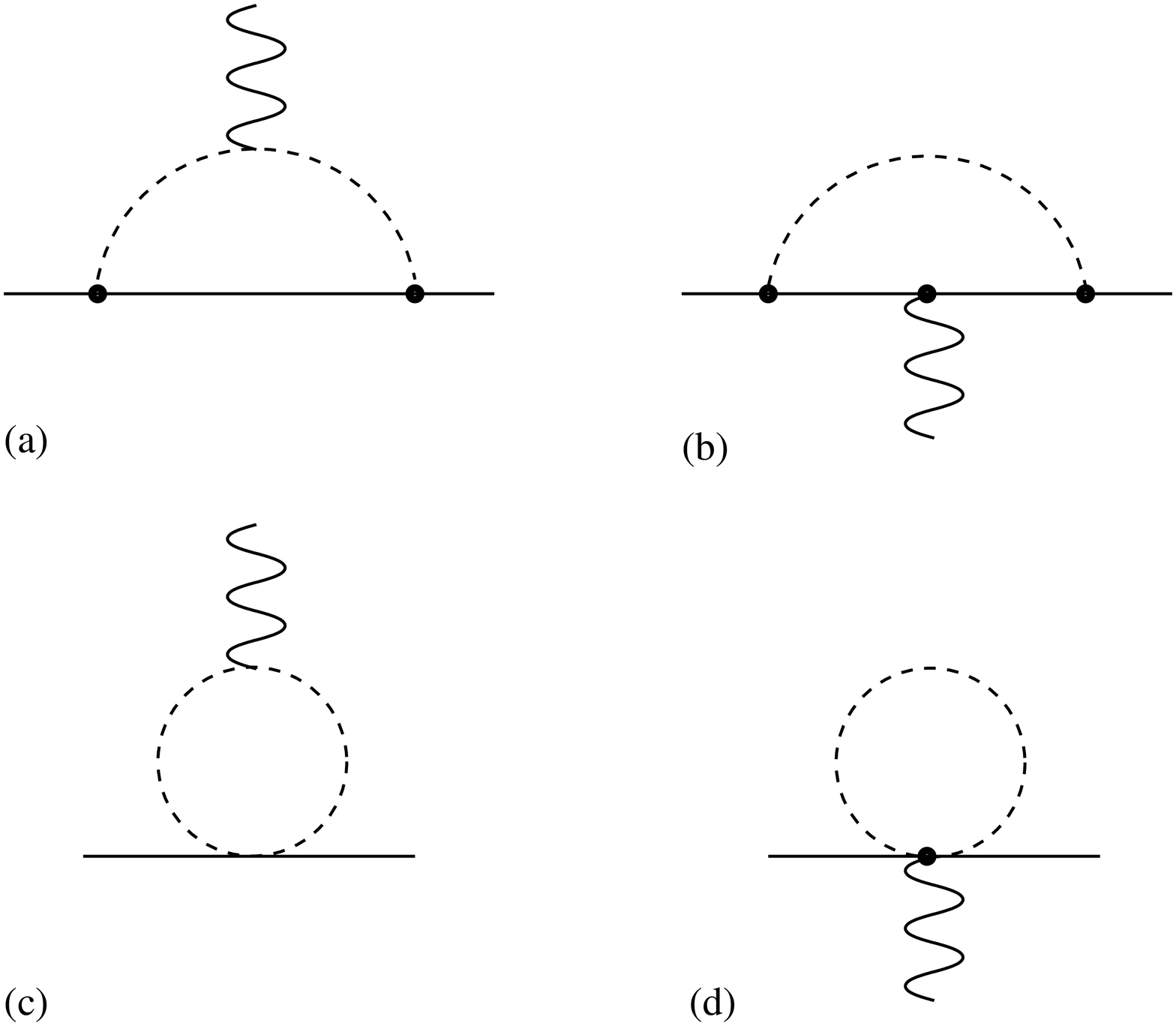}}
\caption{\label{fig:oneloop}Feynman diagrams that yield one-loop corrections to the Dirac form factor. Dashed lines and solid lines denote mesons and baryons, respectively. The inner solid lines in (a) and (b) can also denote decuplet baryons. Although the wave-function renormalization graphs are not displayed, they nevertheless have been included in the analysis.}
\end{figure}

At one-loop level, the Dirac form factor will get corrections arising from the Feynman graphs depicted in Fig.~\ref{fig:oneloop}. These graphs can be written as the product of a baryon operator times a flavor tensor that comprises the loop integrals, which in turn contain the full dependence on $q^2$. Thus,
\begin{equation}
F_{1,\mathrm{Loop}} = \langle \delta O_{\textrm{(a)}} \rangle + \langle \delta O_{\textrm{(b)}} \rangle + \langle \delta O_{\textrm{(c)}} \rangle + \langle \delta O_{\textrm{(d)}} \rangle, \label{eq:ffe}
\end{equation}
where $\delta O_{\textrm{(i)}}$ denotes the correction emerging from diagram \textrm(i) of Fig.~\ref{fig:oneloop}. These operator structures have been analyzed in detail in Ref.~\cite{rfm14b}, where the notation, conventions, and full details of the formalism are provided. A substantial amount of work will be saved by borrowing those operator structures that are applicable and adaptable to the present case; the remaining ones will be analyzed accordingly. Nevertheless, the approach used in Ref.~\cite{rfm14b} will be the starting point. The individual contributions are described in the following section.

\subsection{Diagram \ref{fig:oneloop}(a)}

The one-loop correction \ref{fig:oneloop}(a) for the Dirac form factor can be written as
\begin{equation}
\delta O_{\textrm{(a)}} = \sum_{\mathsf{j}} A^{ia} \mathcal{P}_{\mathsf{j}} A^{ib} P^{ab}(\Delta_{\mathsf{j}}), \label{eq:loop1a}
\end{equation}
where the axial vector current operators $A^{ia}$ and $A^{jb}$ are used at the meson-baryon vertices, $\mathcal{P}_{\mathsf{j}}$ is the spin projection operator for spin $J=\mathsf{j}$ \cite{jen96} so the baryon propagator turns out to be
\begin{equation}
\frac{i\mathcal{P}_{\mathsf{j}}}{k^0-\Delta_{\mathsf{j}}}, \label{eq:barprop}
\end{equation}
and $\Delta_{\mathsf{j}}$ stands for the difference of the hyperfine mass splitting between the intermediate baryon with spin $J=\mathsf{j}$ and the external baryon, specifically,
\begin{equation}
\Delta_{\mathsf{j}} = \mathcal{M}_{\textrm{hyperfine}}|_{J^2=\mathsf{j}(\mathsf{j}+1)}-\mathcal{M}_{\textrm{hyperfine}}|_{J^2=\mathsf{j}_{\textrm{ext}}(\mathsf{j}_{\textrm{ext}}+1)}.
\end{equation}
In Eq.~(\ref{eq:loop1a}) the sum over spin $\mathsf{j}$ is explicitly indicated whereas the sums over repeated spin and flavor indices of the SU(2)$\times$SU(3) algebra are understood. 

For the lowest-lying baryons, at $N_c=3$, the spin projectors and mass splittings acquire simple forms \cite{jen96}, namely,
\begin{subequations}
\label{eq:projnc3}
\begin{eqnarray}
\mathcal{P}_\frac12 & = & -\frac13 \left(J^2-\frac{15}{4}\right), \\
\mathcal{P}_\frac32 & = & \frac13 \left(J^2-\frac34\right),
\end{eqnarray}
\end{subequations}
and
\begin{subequations}
\begin{equation}
\Delta_\frac12 = \left\{
\begin{array}{ll}
\displaystyle 0, & \mathsf{j}_{\textrm{ext}}=\frac12, \\[2mm]
\displaystyle -\Delta, & \mathsf{j}_{\textrm{ext}}=\frac32,
\end{array}
\right.
\end{equation}
\begin{equation}
\Delta_\frac32 = \left\{
\begin{array}{ll}
\displaystyle \Delta, & \mathsf{j}_{\textrm{ext}}=\frac12, \\[2mm]
\displaystyle 0, & \mathsf{j}_{\textrm{ext}}=\frac32, \\[2mm]
\end{array}
\right.
\end{equation}
\end{subequations}
where $\Delta$ is the decuplet-octet mass difference, $\Delta \equiv M_T-M_B$.

Additionally, $P^{ab}$ is an antisymmetric tensor that comprises the integrals over the loops. It can be decomposed as \cite{dai}
\begin{equation}
P^{ab}(\Delta_{\textsf{j}}) = A_0(\Delta_{\textsf{j}}) i\Gamma_0^{ab} + A_1(\Delta_{\textsf{j}}) i\Gamma_1^{ab} + A_2(\Delta_{\textsf{j}}) i\Gamma_2^{ab},
\end{equation}
where the tensors $\Gamma_i^{ab}$ are given by
\begin{subequations}
\begin{eqnarray}
&  & \Gamma_0^{ab} = f^{abQ}, \\
&  & \Gamma_1^{ab} = f^{ab\overline{Q}}, \\
&  & \Gamma_2^{ab} = f^{aeQ}d^{be8} - f^{beQ}d^{ae8} - f^{abe}d^{eQ8}. \label{eq:tens}
\end{eqnarray}
\end{subequations}
$\Gamma_0^{ab}$ and $\Gamma_1^{ab}$ are both SU(3) octets; the former transform like the electric charge and the latter also transforms like the electric charge rotated by $\pi$ in isospin space. $\Gamma_2^{ab}$, on the other hand, breaks SU(3) as a $\mathbf{10} + \mathbf{\overline{10}}$ \cite{dai}. 

The integral over the loop, $I_a(m,\delta,\mu;q^2)$, which contains the full dependence on the momentum transfer, comes into play through
\begin{subequations}
\label{eq:ais}
\begin{eqnarray}
A_0(\Delta_{\textsf{j}}) & = & \frac13 [ I_a(m_\pi,\Delta_{\textsf{j}},\mu;q^2)+2I_a(m_K,\Delta_{\textsf{j}},\mu;q^2) ], \\
A_1(\Delta_{\textsf{j}}) & = & \frac13 [ I_a(m_\pi,\Delta_{\textsf{j}},\mu;q^2)-I_a(m_K,\Delta_{\textsf{j}},\mu;q^2) ], \\
A_2(\Delta_{\textsf{j}}) & = & -\frac{1}{\sqrt{3}}[ I_a(m_\pi,\Delta_{\textsf{j}},\mu;q^2)-I_a(m_K,\Delta_{\textsf{j}},\mu;q^2) ],
\end{eqnarray}
\end{subequations}
and is given by
\begin{eqnarray}
16\pi^2 F_\pi^2 I_a(m,\delta,\mu;q^2) & = & \left[ m^2-2\delta^2+\frac{1}{18}q^2 \right] \left[ -\lambda_\epsilon -1 + \ln \frac{m^2}{\mu^2} \right] - \frac{16}{9}m^2 + 6\delta^2 - \frac{11}{54} q^2 \nonumber \\
&  & \mbox{} - \frac{16m^2+q^2-36\delta^2}{18q^2} \sqrt{q^2(4m^2+q^2)} \ln \left[ \frac{-q^2+\sqrt{q^2(4m^2+q^2)}}{q^2+\sqrt{q^2(4m^2+q^2)}} \right] \nonumber \\
&  & \mbox{} - \int_0^1 dx \frac{2\delta}{\sqrt{\delta^2-q^2(1-x)x-m^2}} \left[ m^2-\delta^2 + \frac43 q^2(1-x)x \right] \ln \left[ \frac{\delta-\sqrt{\delta^2-q^2(1-x)x-m^2}}{\delta+\sqrt{\delta^2-q^2(1-x)x-m^2}} \right], \nonumber \\
\label{eq:ia}
\end{eqnarray}
where $F_\pi$ is the pion decay constant and $\mu$ is the scale parameter of dimensional regularization.

To proceed further, the term $A^{ia}\mathcal{P}_{\mathsf{j}}A^{ib}$ can be decomposed into $\alpha_1 A^{ia} A^{ib}$ and $\alpha_2 A^{ia} J^2 A^{ib}$, where $\alpha_1$ and $\alpha_2$ are some coefficients. An explicit calculation yields \cite{rfm14b}
\begin{equation}
if^{acb}A^{ia}A^{ib} = \sum_{n=1}^7 a_{n}^{\mathbf{8}} S_n^c, \label{eq:d8}
\end{equation}
and
\begin{equation}
if^{acb}A^{ia} J^2 A^{ib} = \sum_{n=1}^7 \overline{a}_{n}^{\mathbf{8}} S_n^c, \label{eq:d8j2}
\end{equation}
for the $\mathbf{8}$ contribution, and
\begin{equation}
i(f^{aec}d^{be8}-f^{bec}d^{ae8}-f^{abe}d^{ec8}) A^{ia}A^{ib} = \sum_{n=1}^{13} b_{n}^{\mathbf{10}+\overline{\mathbf{10}}} O_n^c, \label{eq:d10}
\end{equation}
and
\begin{equation}
i(f^{aec}d^{be8}-f^{bec}d^{ae8}-f^{abe}d^{ec8}) A^{ia} J^2 A^{ib} = \sum_{n=1}^{13} \overline{b}_{n}^{\mathbf{10}+\overline{\mathbf{10}}} O_n^c, \label{eq:d10j2}
\end{equation}
for the $\mathbf{10}+\overline{\mathbf{10}}$ contribution. For computational purposes, a free flavor index $c$ is left in Eqs.~(\ref{eq:d8})--(\ref{eq:d10j2}), which can be set to $Q$ or $\overline{Q}$, as the case may be.

In the large-$N_c$ limit, the one-body operators $T^3$ and $T^8$ are orders $\mathcal{O}(N_c^0)$ and $\mathcal{O}(N_c)$, respectively. $\delta O_{\textrm(a)}$, being proportional to $(g_1/F_\pi)^2$, is naively expected to be order $\mathcal{O}(\sqrt{N_c})^2$ because $g_1$ and $F_\pi$ are orders $\mathcal{O}(N_c)$ and $\mathcal{O}(\sqrt{N_c})$, respectively. However, a close inspection to the operator structure $A^{ia}\mathcal{P}_{\mathsf{j}}A^{jb}$ indicates that it is order $\mathcal{O}(N_c)$ \cite{rfm14b}, so $\delta O_{\textrm(a)}^c$ is order $\mathcal{O}(N_c^0)$, or equivalently, $1/N_c$ times the tree-level value.

On the other hand, the coefficients $a_{n}^{\mathbf{8}}$, $\overline{a}_{n}^{\mathbf{8}}$, $b_{n}^{\mathbf{10}+\overline{\mathbf{10}}}$, and $\overline{b}_{n}^{\mathbf{10}+\overline{\mathbf{10}}}$ are listed in full in Appendix C of Ref.~\cite{rfm14b} through Eqs.~(C1)--(C4), respectively. It is convenient, however, to reproduce here the operator bases $S_m^c$ and $O_n^c$, namely,
\begin{eqnarray}
\begin{array}{lll}
S_1^c = T^c, &
S_2^c = \{J^r, G^{rc}\}, & 
S_3^c = \{J^2, T^c\}, \\
S_4^c = \{J^2, \{J^r, G^{rc}\}\}, & 
S_5^c = \{J^2, \{J^2, T^c\}\}, &
S_6^c = \{J^2, \{J^2, \{J^r, G^{rc}\}\}\}, \\
S_7^c = \{J^2, \{J^2, \{J^2, T^c\}\}\}, & & \label{eq:1op}
\end{array}
\end{eqnarray}
and
\begin{eqnarray}
\begin{array}{lll}
O_{1}^c = d^{c8e} T^e, &
O_{2}^c =d^{c8e} \{J^r,G^{re}\} , &
O_{3}^c =d^{c8e} \{J^2,T^e\}, \\
O_{4}^c = \{T^c,\{J^r,G^{r8}\}\}, &
O_{5}^c = \{T^8,\{J^r,G^{rc}\}\}, &
O_{6}^c = d^{c8e} \{J^2,\{J^r,G^{re}\}\}, \\
O_{7}^c = d^{c8e} \{J^2,\{J^2,T^e\}\}, &
O_{8}^c = \{J^2,\{T^c,\{J^r,G^{r8}\}\}\}, &
O_{9}^c = \{J^2,\{T^8,\{J^r,G^{rc}\}\}\}, \\
O_{10}^c = d^{c8e} \{J^2,\{J^2,\{J^r,G^{re}\}\}\}, &
O_{11}^c = d^{c8e} \{J^2,\{J^2,\{J^2,T^e\}\}\}, &
O_{12}^c = \{J^2,\{J^2,\{T^c,\{J^r,G^{r8}\}\}\}\}, \\
O_{13}^c = \{J^2,\{J^2,\{T^8,\{J^r,G^{rc}\}\}\}\}. & & \label{eq:8op}
\end{array}
\end{eqnarray}
The matrix elements of the operator bases (\ref{eq:1op}) and (\ref{eq:8op}) are listed in Tables \ref{t:1opQbo}--\ref{t:8opQbd} for octet and decuplet baryons, for both flavor indices $c=3$ and $c=8$.

\begin{table*}
\caption{\label{t:1opQbo}Matrix elements of singlet operators between baryon octet states.}
\begin{ruledtabular}
\begin{tabular}{cccccccccc}
& $n$ & $p$ & $ \Sigma^-$ & $ \Sigma^0$ & $ \Sigma^+$ & $ \Xi^-$ & $ \Xi^0$ & $ \Lambda$ & $ \Lambda\Sigma^0$ \\
\hline
$ \langle S_{1}^3 \rangle$ & $-\frac12$ & $\frac12$ & $-1$ & $0$ & $1$ & $-\frac12$ & $\frac12$ & $0$ & $0$ \\
$ \langle S_{2}^3 \rangle$ & $-\frac54$ & $\frac54$ & $-1$ & $0$ & $1$ & $\frac14$ & $-\frac14$ & $0$ & $\frac{\sqrt{3}}{2}$ \\
$ \langle S_{3}^3 \rangle$ & $-\frac34$ & $\frac34$ & $-\frac{3}{2}$ & $0$ & $\frac{3}{2}$ & $-\frac34$ & $\frac34$ & $0$ & $0$ \\
$ \langle S_{4}^3 \rangle$ & $-\frac{15}{8}$ & $\frac{15}{8}$ & $-\frac{3}{2}$ & $0$ & $\frac{3}{2}$ & $\frac38$ & $-\frac38$ & $0$ & $\frac{3 \sqrt{3}}{4}$ \\
$ \langle S_{5}^3 \rangle$ & $-\frac98$ & $\frac98$ & $-\frac{9}{4}$ & $0$ & $\frac{9}{4}$ & $-\frac98$ & $\frac98$ & $0$ & $0$ \\
$ \langle S_{6}^3 \rangle$ & $-\frac{45}{16}$ & $\frac{45}{16}$ & $-\frac{9}{4}$ & $0$ & $\frac{9}{4}$ & $\frac{9}{16}$ & $-\frac{9}{16}$ & $0$ & $\frac{9 \sqrt{3}}{8}$ \\
$ \langle S_{7}^3 \rangle$ & $-\frac{27}{16}$ & $\frac{27}{16}$ & $-\frac{27}{8}$ & $0$ & $\frac{27}{8}$ & $-\frac{27}{16}$ & $\frac{27}{16}$ & $0$ & $0$ \\
$ \langle S_{1}^8 \rangle$ & $\frac{\sqrt{3}}{2}$ & $\frac{\sqrt{3}}{2}$ & $0$ & $0$ & $0$ & $-\frac{\sqrt{3}}{2}$ & $-\frac{\sqrt{3}}{2}$ & $0$ & $0$ \\
$ \langle S_{2}^8 \rangle$ & $\frac{\sqrt{3}}{4}$ & $\frac{\sqrt{3}}{4}$ & $\frac{\sqrt{3}}{2}$ & $\frac{\sqrt{3}}{2}$ & $\frac{\sqrt{3}}{2}$ & $-\frac{3 \sqrt{3}}{4}$ & $-\frac{3 \sqrt{3}}{4}$ & $-\frac{\sqrt{3}}{2}$ & $0$ \\
$ \langle S_{3}^8 \rangle$ & $\frac{3 \sqrt{3}}{4}$ & $\frac{3 \sqrt{3}}{4}$ & $0$ & $0$ & $0$ & $-\frac{3 \sqrt{3}}{4}$ & $-\frac{3 \sqrt{3}}{4}$ & $0$ & $0$ \\
$ \langle S_{4}^8 \rangle$ & $\frac{3 \sqrt{3}}{8}$ & $\frac{3 \sqrt{3}}{8}$ & $\frac{3 \sqrt{3}}{4}$ & $\frac{3 \sqrt{3}}{4}$ & $\frac{3 \sqrt{3}}{4}$ & $-\frac{9 \sqrt{3}}{8}$ & $-\frac{9 \sqrt{3}}{8}$ & $-\frac{3 \sqrt{3}}{4}$ & $0$ \\
$ \langle S_{5}^8 \rangle$ & $\frac{9 \sqrt{3}}{8}$ & $\frac{9 \sqrt{3}}{8}$ & $0$ & $0$ & $0$ & $-\frac{9 \sqrt{3}}{8}$ & $-\frac{9 \sqrt{3}}{8}$ & $0$ & $0$ \\
$ \langle S_{6}^8 \rangle$ & $\frac{9 \sqrt{3}}{16}$ & $\frac{9 \sqrt{3}}{16}$ & $\frac{9 \sqrt{3}}{8}$ & $\frac{9 \sqrt{3}}{8}$ & $\frac{9 \sqrt{3}}{8}$ & $-\frac{27 \sqrt{3}}{16}$ & $-\frac{27 \sqrt{3}}{16}$ & $-\frac{9 \sqrt{3}}{8}$ & $0$ \\
$ \langle S_{7}^8 \rangle$ & $\frac{27 \sqrt{3}}{16}$ & $\frac{27 \sqrt{3}}{16}$ & $0$ & $0$ & $0$ & $-\frac{27 \sqrt{3}}{16}$ & $-\frac{27 \sqrt{3}}{16}$ & $0$ & $0$
\end{tabular}
\end{ruledtabular}
\end{table*}

\begin{table*}
\caption{\label{t:1opQbd}Matrix elements of singlet operators between baryon decuplet states.}
\begin{ruledtabular}
\begin{tabular}{ccccccccccc}
& $\Delta^{++}$ & $\Delta^+$ & $\Delta^0$ & $\Delta^-$ & ${\Sigma^*}^+$ & ${\Sigma^*}^0$ & ${\Sigma^*}^-$ & ${\Xi^*}^0$ & ${\Xi^*}^-$ & $\Omega^-$ \\
\hline
$\langle S_{1}^3 \rangle$ & $\frac{3}{2}$ & $\frac12$ & $-\frac12$ & $-\frac{3}{2}$ & $1$ & $0$ & $-1$ & $\frac12$ & $-\frac12$ & $0$ \\
$\langle S_{2}^3 \rangle$ & $\frac{15}{4}$ & $\frac54$ & $-\frac54$ & $-\frac{15}{4}$ & $\frac{5}{2}$ & $0$ & $-\frac{5}{2}$ & $\frac54$ & $-\frac54$ & $0$ \\
$\langle S_{3}^3 \rangle$ & $\frac{45}{4}$ & $\frac{15}{4}$ & $-\frac{15}{4}$ & $-\frac{45}{4}$ & $\frac{15}{2}$ & $0$ & $-\frac{15}{2}$ & $\frac{15}{4}$ & $-\frac{15}{4}$ & $0$ \\
$\langle S_{4}^3 \rangle$ & $\frac{225}{8}$ & $\frac{75}{8}$ & $-\frac{75}{8}$ & $-\frac{225}{8}$ & $\frac{75}{4}$ & $0$ & $-\frac{75}{4}$ & $\frac{75}{8}$ & $-\frac{75}{8}$ & $0$ \\
$\langle S_{5}^3 \rangle$ & $\frac{675}{8}$ & $\frac{225}{8}$ & $-\frac{225}{8}$ & $-\frac{675}{8}$ & $\frac{225}{4}$ & $0$ & $-\frac{225}{4}$ & $\frac{225}{8}$ & $-\frac{225}{8}$ & $0$ \\
$\langle S_{6}^3 \rangle$ & $\frac{3375}{16}$ & $\frac{1125}{16}$ & $-\frac{1125}{16}$ & $-\frac{3375}{16}$ & $\frac{1125}{8}$ & $0$ & $-\frac{1125}{8}$ & $\frac{1125}{16}$ & $-\frac{1125}{16}$ & $0$ \\
$\langle S_{7}^3 \rangle$ & $\frac{10125}{16}$ & $\frac{3375}{16}$ & $-\frac{3375}{16}$ & $-\frac{10125}{16}$ & $\frac{3375}{8}$ & $0$ & $-\frac{3375}{8}$ & $\frac{3375}{16}$ & $-\frac{3375}{16}$ & $0$ \\
$\langle S_{1}^8 \rangle$ & $\frac{\sqrt{3}}{2}$ & $\frac{\sqrt{3}}{2}$ & $\frac{\sqrt{3}}{2}$ & $\frac{\sqrt{3}}{2}$ & $0$ & $0$ & $0$ & $-\frac{\sqrt{3}}{2}$ & $-\frac{\sqrt{3}}{2}$ & $-\sqrt{3}$ \\
$\langle S_{2}^8 \rangle$ & $\frac{5 \sqrt{3}}{4}$ & $\frac{5 \sqrt{3}}{4}$ & $\frac{5 \sqrt{3}}{4}$ & $\frac{5 \sqrt{3}}{4}$ & $0$ & $0$ & $0$ & $-\frac{5 \sqrt{3}}{4}$ & $-\frac{5 \sqrt{3}}{4}$ & $-\frac{5 \sqrt{3}}{2}$ \\
$\langle S_{3}^8 \rangle$ & $\frac{15 \sqrt{3}}{4}$ & $\frac{15 \sqrt{3}}{4}$ & $\frac{15 \sqrt{3}}{4}$ & $\frac{15 \sqrt{3}}{4}$ & $0$ & $0$ & $0$ & $-\frac{15 \sqrt{3}}{4}$ & $-\frac{15 \sqrt{3}}{4}$ & $-\frac{15 \sqrt{3}}{2}$ \\
$\langle S_{4}^8 \rangle$ & $\frac{75 \sqrt{3}}{8}$ & $\frac{75 \sqrt{3}}{8}$ & $\frac{75 \sqrt{3}}{8}$ & $\frac{75 \sqrt{3}}{8}$ & $0$ & $0$ & $0$ & $-\frac{75 \sqrt{3}}{8}$ & $-\frac{75 \sqrt{3}}{8}$ & $-\frac{75 \sqrt{3}}{4}$ \\
$\langle S_{5}^8 \rangle$ & $\frac{225 \sqrt{3}}{8}$ & $\frac{225 \sqrt{3}}{8}$ & $\frac{225 \sqrt{3}}{8}$ & $\frac{225 \sqrt{3}}{8}$ & $0$ & $0$ & $0$ & $-\frac{225 \sqrt{3}}{8}$ & $-\frac{225\sqrt{3}}{8}$ & $-\frac{225 \sqrt{3}}{4}$ \\
$\langle S_{6}^8 \rangle$ & $\frac{1125 \sqrt{3}}{16}$ & $\frac{1125 \sqrt{3}}{16}$ & $\frac{1125 \sqrt{3}}{16}$ & $\frac{1125 \sqrt{3}}{16}$ & $0$ & $0$ & $0$ & $-\frac{1125 \sqrt{3}}{16}$ & $-\frac{1125 \sqrt{3}}{16}$ & $-\frac{1125 \sqrt{3}}{8}$ \\
$\langle S_{7}^8 \rangle$ & $\frac{3375 \sqrt{3}}{16}$ & $\frac{3375 \sqrt{3}}{16}$ & $\frac{3375 \sqrt{3}}{16}$ & $\frac{3375 \sqrt{3}}{16}$ & $0$ & $0$ & $0$ & $-\frac{3375 \sqrt{3}}{16}$ & $-\frac{3375 \sqrt{3}}{16}$ & $-\frac{3375 \sqrt{3}}{8}$
\end{tabular}
\end{ruledtabular}
\end{table*}

\begin{table*}
\caption{\label{t:8opQbo}Matrix elements of octet operators between baryon octet states.}
\begin{ruledtabular}
\begin{tabular}{cccccccccc}
& $n$ & $p$ & $ \Sigma^-$ & $ \Sigma^0$ & $ \Sigma^+$ & $ \Xi^-$ & $ \Xi^0$ & $ \Lambda$ & $ \Lambda\Sigma^0$ \\
\hline
$\langle O_{1}^3 \rangle$ & $-\frac{1}{2 \sqrt{3}}$ & $\frac{1}{2 \sqrt{3}}$ & $-\frac{1}{\sqrt{3}}$ & $0$ & $\frac{1}{\sqrt{3}}$ & $-\frac{1}{2 \sqrt{3}}$ & $\frac{1}{2 \sqrt{3}}$ & $0$ & $0$ \\
$\langle O_{2}^3 \rangle$ & $-\frac{5}{4 \sqrt{3}}$ & $\frac{5}{4 \sqrt{3}}$ & $-\frac{1}{\sqrt{3}}$ & $0$ & $\frac{1}{\sqrt{3}}$ & $\frac{1}{4 \sqrt{3}}$ & $-\frac{1}{4 \sqrt{3}}$ & $0$ & $\frac12$ \\
$\langle O_{3}^3 \rangle$ & $-\frac{\sqrt{3}}{4}$ & $\frac{\sqrt{3}}{4}$ & $-\frac{\sqrt{3}}{2}$ & $0$ & $\frac{\sqrt{3}}{2}$ & $-\frac{\sqrt{3}}{4}$ & $\frac{\sqrt{3}}{4}$ & $0$ & $0$ \\
$\langle O_{4}^3 \rangle$ & $-\frac{\sqrt{3}}{4}$ & $\frac{\sqrt{3}}{4}$ & $-\sqrt{3}$ & $0$ & $ \sqrt{3}$ & $\frac{3 \sqrt{3}}{4}$ & $-\frac{3 \sqrt{3}}{4}$ & $0$ & $0$ \\
$\langle O_{5}^3 \rangle$ & $-\frac{5 \sqrt{3}}{4}$ & $\frac{5 \sqrt{3}}{4}$ & $0$ & $0$ & $0$ & $-\frac{\sqrt{3}}{4}$ & $\frac{\sqrt{3}}{4}$ & $0$ & $0$ \\
$\langle O_{6}^3 \rangle$ & $-\frac{5 \sqrt{3}}{8}$ & $\frac{5 \sqrt{3}}{8}$ & $-\frac{\sqrt{3}}{2}$ & $0$ & $\frac{\sqrt{3}}{2}$ & $\frac{\sqrt{3}}{8}$ & $-\frac{\sqrt{3}}{8}$ & $0$ & $\frac34$ \\
$\langle O_{7}^3 \rangle$ & $-\frac{3 \sqrt{3}}{8}$ & $\frac{3 \sqrt{3}}{8}$ & $-\frac{3 \sqrt{3}}{4}$ & $0$ & $\frac{3 \sqrt{3}}{4}$ & $-\frac{3 \sqrt{3}}{8}$ & $\frac{3 \sqrt{3}}{8}$ & $0$ & $0$ \\
$\langle O_{8}^3 \rangle$ & $-\frac{3 \sqrt{3}}{8}$ & $\frac{3 \sqrt{3}}{8}$ & $-\frac{3 \sqrt{3}}{2}$ & $0$ & $\frac{3 \sqrt{3}}{2}$ & $\frac{9 \sqrt{3}}{8}$ & $-\frac{9 \sqrt{3}}{8}$ & $0$ & $0$ \\
$\langle O_{9}^3 \rangle$ & $-\frac{15 \sqrt{3}}{8}$ & $\frac{15 \sqrt{3}}{8}$ & $0$ & $0$ & $0$ & $-\frac{3 \sqrt{3}}{8}$ & $\frac{3 \sqrt{3}}{8}$ & $0$ & $0$ \\
$\langle O_{10}^3 \rangle$ & $-\frac{15 \sqrt{3}}{16}$ & $\frac{15 \sqrt{3}}{16}$ & $-\frac{3 \sqrt{3}}{4}$ & $0$ & $\frac{3 \sqrt{3}}{4}$ & $\frac{3 \sqrt{3}}{16}$ & $-\frac{3 \sqrt{3}}{16}$ & $0$ & $\frac98$ \\
$\langle O_{11}^3 \rangle$ & $-\frac{9 \sqrt{3}}{16}$ & $\frac{9 \sqrt{3}}{16}$ & $-\frac{9 \sqrt{3}}{8}$ & $0$ & $\frac{9 \sqrt{3}}{8}$ & $-\frac{9 \sqrt{3}}{16}$ & $ \frac{9 \sqrt{3}}{16}$ & $0$ & $0$ \\
$\langle O_{12}^3 \rangle$ & $-\frac{9 \sqrt{3}}{16}$ & $\frac{9 \sqrt{3}}{16}$ & $-\frac{9 \sqrt{3}}{4}$ & $0$ & $\frac{9 \sqrt{3}}{4}$ & $\frac{27 \sqrt{3}}{16}$ & $-\frac{27 \sqrt{3}}{16}$ & $0$ & $0$ \\
$\langle O_{13}^3 \rangle$ & $-\frac{45 \sqrt{3}}{16}$ & $\frac{45 \sqrt{3}}{16}$ & $0$ & $0$ & $0$ & $-\frac{9 \sqrt{3}}{16}$ & $\frac{9 \sqrt{3}}{16}$ & $0$ & $0$ \\
$\langle O_{1}^8 \rangle$ & $-\frac12$ & $-\frac12$ & $0$ & $0$ & $0$ & $\frac12$ & $\frac12$ & $0$ & $0$ \\
$\langle O_{2}^8 \rangle$ & $-\frac14$ & $-\frac14$ & $-\frac12$ & $-\frac12$ & $-\frac12$ & $\frac34$ & $\frac34$ & $\frac12$ & $0$ \\
$\langle O_{3}^8 \rangle$ & $-\frac34$ & $-\frac34$ & $0$ & $0$ & $0$ & $\frac34$ & $\frac34$ & $0$ & $0$ \\
$\langle O_{4}^8 \rangle$ & $\frac34$ & $\frac34$ & $0$ & $0$ & $0$ & $\frac{9}{4}$ & $\frac{9}{4}$ & $0$ & $0$ \\
$\langle O_{5}^8 \rangle$ & $\frac34$ & $\frac34$ & $0$ & $0$ & $0$ & $\frac{9}{4}$ & $\frac{9}{4}$ & $0$ & $0$ \\
$\langle O_{6}^8 \rangle$ & $-\frac38$ & $-\frac38$ & $-\frac34$ & $-\frac34$ & $-\frac34$ & $\frac98$ & $\frac98$ & $\frac34$ & $0$ \\
$\langle O_{7}^8 \rangle$ & $-\frac98$ & $-\frac98$ & $0$ & $0$ & $0$ & $\frac98$ & $\frac98$ & $0$ & $0$ \\
$\langle O_{8}^8 \rangle$ & $\frac98$ & $\frac98$ & $0$ & $0$ & $0$ & $\frac{27}{8}$ & $\frac{27}{8}$ & $0$ & $0$ \\
$\langle O_{9}^8 \rangle$ & $\frac98$ & $\frac98$ & $0$ & $0$ & $0$ & $\frac{27}{8}$ & $\frac{27}{8}$ & $0$ & $0$ \\
$\langle O_{10}^8 \rangle$ & $-\frac{9}{16}$ & $-\frac{9}{16}$ & $-\frac98$ & $-\frac98$ & $-\frac98$ & $\frac{27}{16}$ & $\frac{27}{16}$ & $\frac98$ & $0$ \\
$\langle O_{11}^8 \rangle$ & $-\frac{27}{16}$ & $-\frac{27}{16}$ & $0$ & $0$ & $0$ & $\frac{27}{16}$ & $\frac{27}{16}$ & $0$ & $0$ \\
$\langle O_{12}^8 \rangle$ & $\frac{27}{16}$ & $\frac{27}{16}$ & $0$ & $0$ & $0$ & $\frac{81}{16}$ & $\frac{81}{16}$ & $0$ & $0$ \\
$\langle O_{13}^8 \rangle$ & $\frac{27}{16}$ & $\frac{27}{16}$ & $0$ & $0$ & $0$ & $\frac{81}{16}$ & $\frac{81}{16}$ & $0$ & $0$
\end{tabular}
\end{ruledtabular}
\end{table*}

\begin{table*}
\caption{\label{t:8opQbd}Matrix elements of octet operators between baryon decuplet states.}
\begin{ruledtabular}
\begin{tabular}{ccccccccccc}
& $\Delta^{++}$ & $\Delta^+$ & $\Delta^0$ & $\Delta^-$ & ${\Sigma^*}^+$ & ${\Sigma^*}^0$ & ${\Sigma^*}^-$ & ${\Xi^*}^0$ & ${\Xi^*}^-$ & $\Omega^-$ \\
\hline
$\langle O_{1}^3 \rangle$ & $\frac{\sqrt{3}}{2}$ & $\frac{1}{2 \sqrt{3}}$ & $-\frac{1}{2 \sqrt{3}}$ & $-\frac{\sqrt{3}}{2}$ & $\frac{1}{\sqrt{3}}$ & $0$ & $-\frac{1}{\sqrt{3}}$ & $\frac{1}{2\sqrt{3}}$ & $-\frac{1}{2 \sqrt{3}}$ & $0$ \\
$\langle O_{2}^3 \rangle$ & $\frac{5 \sqrt{3}}{4}$ & $\frac{5}{4 \sqrt{3}}$ & $-\frac{5}{4 \sqrt{3}}$ & $-\frac{5 \sqrt{3}}{4}$ & $\frac{5}{2 \sqrt{3}}$ & $0$ & $-\frac{5}{2 \sqrt{3}}$ & $\frac{5}{4 \sqrt{3}}$ & $-\frac{5}{4 \sqrt{3}}$ & $0$ \\
$\langle O_{3}^3 \rangle$ & $\frac{15 \sqrt{3}}{4}$ & $\frac{5 \sqrt{3}}{4}$ & $-\frac{5 \sqrt{3}}{4}$ & $-\frac{15 \sqrt{3}}{4}$ & $\frac{5 \sqrt{3}}{2}$ & $0$ & $-\frac{5 \sqrt{3}}{2}$ & $\frac{5 \sqrt{3}}{4}$ & $-\frac{5 \sqrt{3}}{4}$ & $0$ \\
$\langle O_{4}^3 \rangle$ & $\frac{15 \sqrt{3}}{4}$ & $\frac{5 \sqrt{3}}{4}$ & $-\frac{5 \sqrt{3}}{4}$ & $-\frac{15 \sqrt{3}}{4}$ & $0$ & $0$ & $0$ & $-\frac{5 \sqrt{3}}{4}$ & $\frac{5 \sqrt{3}}{4}$ & $0$ \\
$\langle O_{5}^3 \rangle$ & $\frac{15 \sqrt{3}}{4}$ & $\frac{5 \sqrt{3}}{4}$ & $-\frac{5 \sqrt{3}}{4}$ & $-\frac{15 \sqrt{3}}{4}$ & $0$ & $0$ & $0$ & $-\frac{5 \sqrt{3}}{4}$ & $\frac{5 \sqrt{3}}{4}$ & $0$ \\
$\langle O_{6}^3 \rangle$ & $\frac{75 \sqrt{3}}{8}$ & $\frac{25 \sqrt{3}}{8}$ & $-\frac{25 \sqrt{3}}{8}$ & $-\frac{75 \sqrt{3}}{8}$ & $\frac{25 \sqrt{3}}{4}$ & $0$ & $-\frac{25 \sqrt{3}}{4}$ & $\frac{25 \sqrt{3}}{8}$ & $-\frac{25 \sqrt{3}}{8}$ & $0$ \\
$\langle O_{7}^3 \rangle$ & $\frac{225 \sqrt{3}}{8}$ & $\frac{75 \sqrt{3}}{8}$ & $-\frac{75 \sqrt{3}}{8}$ & $-\frac{225 \sqrt{3}}{8}$ & $\frac{75 \sqrt{3}}{4}$ & $0$ & $-\frac{75 \sqrt{3}}{4}$ & $\frac{75 \sqrt{3}}{8}$ & $-\frac{75 \sqrt{3}}{8}$ & $0$ \\
$\langle O_{8}^3 \rangle$ & $\frac{225 \sqrt{3}}{8}$ & $\frac{75 \sqrt{3}}{8}$ & $-\frac{75 \sqrt{3}}{8}$ & $-\frac{225 \sqrt{3}}{8}$ & $0$ & $0$ & $0$ & $-\frac{75 \sqrt{3}}{8}$ & $\frac{75\sqrt{3}}{8}$ & $0$ \\
$\langle O_{9}^3 \rangle$ & $\frac{225 \sqrt{3}}{8}$ & $\frac{75 \sqrt{3}}{8}$ & $-\frac{75 \sqrt{3}}{8}$ & $-\frac{225 \sqrt{3}}{8}$ & $0$ & $0$ & $0$ & $-\frac{75 \sqrt{3}}{8}$ & $\frac{75\sqrt{3}}{8}$ & $0$ \\
$\langle O_{10}^3 \rangle$ & $\frac{1125 \sqrt{3}}{16}$ & $\frac{375 \sqrt{3}}{16}$ & $-\frac{375 \sqrt{3}}{16}$ & $-\frac{1125 \sqrt{3}}{16}$ & $\frac{375 \sqrt{3}}{8}$ & $0$ & $-\frac{375\sqrt{3}}{8}$ & $\frac{375 \sqrt{3}}{16}$ & $-\frac{375 \sqrt{3}}{16}$ & $0$ \\
$\langle O_{11}^3 \rangle$ & $\frac{3375 \sqrt{3}}{16}$ & $\frac{1125 \sqrt{3}}{16}$ & $-\frac{1125 \sqrt{3}}{16}$ & $-\frac{3375 \sqrt{3}}{16}$ & $\frac{1125 \sqrt{3}}{8}$ & $0$ & $-\frac{1125 \sqrt{3}}{8}$ & $\frac{1125 \sqrt{3}}{16}$ & $-\frac{1125 \sqrt{3}}{16}$ & $0$ \\
$\langle O_{12}^3 \rangle$ & $\frac{3375 \sqrt{3}}{16}$ & $\frac{1125 \sqrt{3}}{16}$ & $-\frac{1125 \sqrt{3}}{16}$ & $-\frac{3375 \sqrt{3}}{16}$ & $0$ & $0$ & $0$ & $-\frac{1125 \sqrt{3}}{16}$ & $\frac{1125 \sqrt{3}}{16}$ & $0$ \\
$\langle O_{13}^3 \rangle$ & $\frac{3375 \sqrt{3}}{16}$ & $\frac{1125 \sqrt{3}}{16}$ & $-\frac{1125 \sqrt{3}}{16}$ & $-\frac{3375 \sqrt{3}}{16}$ & $0$ & $0$ & $0$ & $-\frac{1125 \sqrt{3}}{16}$ & $\frac{1125 \sqrt{3}}{16}$ & $0$ \\
$\langle O_{1}^8 \rangle$ & $-\frac12$ & $-\frac12$ & $-\frac12$ & $-\frac12$ & $0$ & $0$ & $0$ & $\frac12$ & $\frac12$ & $1$ \\
$\langle O_{2}^8 \rangle$ & $-\frac54$ & $-\frac54$ & $-\frac54$ & $-\frac54$ & $0$ & $0$ & $0$ & $\frac54$ & $\frac54$ & $\frac{5}{2}$ \\
$\langle O_{3}^8 \rangle$ & $-\frac{15}{4}$ & $-\frac{15}{4}$ & $-\frac{15}{4}$ & $-\frac{15}{4}$ & $0$ & $0$ & $0$ & $\frac{15}{4}$ & $\frac{15}{4}$ & $\frac{15}{2}$ \\
$\langle O_{4}^8 \rangle$ & $\frac{15}{4}$ & $\frac{15}{4}$ & $\frac{15}{4}$ & $\frac{15}{4}$ & $0$ & $0$ & $0$ & $\frac{15}{4}$ & $\frac{15}{4}$ & $15$ \\
$\langle O_{5}^8 \rangle$ & $\frac{15}{4}$ & $\frac{15}{4}$ & $\frac{15}{4}$ & $\frac{15}{4}$ & $0$ & $0$ & $0$ & $\frac{15}{4}$ & $\frac{15}{4}$ & $15$ \\
$\langle O_{6}^8 \rangle$ & $-\frac{75}{8}$ & $-\frac{75}{8}$ & $-\frac{75}{8}$ & $-\frac{75}{8}$ & $0$ & $0$ & $0$ & $\frac{75}{8}$ & $\frac{75}{8}$ & $\frac{75}{4}$ \\
$\langle O_{7}^8 \rangle$ & $-\frac{225}{8}$ & $-\frac{225}{8}$ & $-\frac{225}{8}$ & $-\frac{225}{8}$ & $0$ & $0$ & $0$ & $\frac{225}{8}$ & $\frac{225}{8}$ & $\frac{225}{4}$ \\
$\langle O_{8}^8 \rangle$ & $\frac{225}{8}$ & $\frac{225}{8}$ & $\frac{225}{8}$ & $\frac{225}{8}$ & $0$ & $0$ & $0$ & $\frac{225}{8}$ & $\frac{225}{8}$ & $\frac{225}{2}$ \\
$\langle O_{9}^8 \rangle$ & $\frac{225}{8}$ & $\frac{225}{8}$ & $\frac{225}{8}$ & $\frac{225}{8}$ & $0$ & $0$ & $0$ & $\frac{225}{8}$ & $\frac{225}{8}$ & $\frac{225}{2}$ \\
$\langle O_{10}^8 \rangle$ & $-\frac{1125}{16}$ & $-\frac{1125}{16}$ & $-\frac{1125}{16}$ & $-\frac{1125}{16}$ & $0$ & $0$ & $0$ & $\frac{1125}{16}$ & $\frac{1125}{16}$ & $\frac{1125}{8}$ \\
$\langle O_{11}^8 \rangle$ & $-\frac{3375}{16}$ & $-\frac{3375}{16}$ & $-\frac{3375}{16}$ & $-\frac{3375}{16}$ & $0$ & $0$ & $0$ & $\frac{3375}{16}$ & $\frac{3375}{16}$ & $\frac{3375}{8}$ \\
$\langle O_{12}^8 \rangle$ & $\frac{3375}{16}$ & $\frac{3375}{16}$ & $\frac{3375}{16}$ & $\frac{3375}{16}$ & $0$ & $0$ & $0$ & $\frac{3375}{16}$ & $\frac{3375}{16}$ & $\frac{3375}{4}$ \\
$\langle O_{13}^8 \rangle$ & $\frac{3375}{16}$ & $\frac{3375}{16}$ & $\frac{3375}{16}$ & $\frac{3375}{16}$ & $0$ & $0$ & $0$ & $\frac{3375}{16}$ & $\frac{3375}{16}$ & $\frac{3375}{4}$ \\
\end{tabular}
\end{ruledtabular}
\end{table*}

Gathering together partial results, the correction $\delta O_{\textrm{(a)}}$, Eq.~(\ref{eq:loop1a}), to the SU(3) symmetric value of the baryon Dirac form factor can be organized as
\begin{eqnarray}
\delta O_{(a)} & = &
\mathcal{P}_{1/2} A^{ia}\mathcal{P}_{1/2}A^{ib}\left[A_0(0)i\Gamma_0^{ab}+A_1(0)i\Gamma_1^{ab}+A_2(0)i\Gamma_2^{ab} \right]\mathcal{P}_{1/2} \nonumber \\
&  & \mbox{} + \mathcal{P}_{1/2} A^{ia}\mathcal{P}_{3/2}A^{ib}\left[A_0(\Delta)i\Gamma_0^{ab}+A_1(\Delta)i\Gamma_1^{ab}+A_2(\Delta)i\Gamma_2^{ab} \right]\mathcal{P}_{1/2}, \label{eq:mm12}
\end{eqnarray}
for octet baryons and
\begin{eqnarray}
\delta O_{(a)} & = &
\mathcal{P}_{3/2} A^{ia}\mathcal{P}_{1/2}A^{ib}\left[A_0(-\Delta)i\Gamma_0^{ab}+A_1(-\Delta)i\Gamma_1^{ab}+A_2(-\Delta)i\Gamma_2^{ab} \right]\mathcal{P}_{3/2} \nonumber \\
&  & \mbox{} + \mathcal{P}_{3/2} A^{ia}\mathcal{P}_{3/2}A^{ib}\left[A_0(0)i\Gamma_0^{ab}+A_1(0)i\Gamma_1^{ab}+A_2(0)i\Gamma_2^{ab} \right]\mathcal{P}_{3/2}, \label{eq:mm32}
\end{eqnarray}
for decuplet baryons.

The matrix element of the operator $\delta O_\mathrm{(a)}$ can now be straightforwardly constructed for any baryon as the sum of products of three factors: an operator coefficient times the corresponding operator matrix element---which can be read off from Tables \ref{t:1opQbo} through \ref{t:8opQbd}---times the corresponding loop integral. As a practical example, the expression for the proton is
\begin{eqnarray}
\langle p|\delta O_\mathrm{(a)}|p\rangle & = & \left[ \frac{25}{24} a_1^2 + \frac{5}{12} a_1b_2 + \frac{25}{36} a_1b_3 + \frac{1}{24} b_2^2 + \frac{5}{36} b_2b_3 + \frac{25}{216} b_3^2 \right] I_a(m_\pi,0,\mu;q^2) \nonumber \\
&  & \mbox{} + \left[ \frac{7}{12} a_1^2 + \frac13 a_1b_2 + \frac{7}{18} a_1b_3 + \frac{1}{12} b_2^2 + \frac19 b_2b_3 + \frac{7}{108} b_3^2 \right] I_a(m_K,0,\mu;q^2) \nonumber \\
&  & \mbox{} + \left[ -\frac23 a_1^2 - \frac23 a_1c_3 - \frac16 c_3^2 \right] I_a(m_\pi,\Delta,\mu;q^2)
+ \left[ \frac16 a_1^2 + \frac16 a_1c_3 + \frac{1}{24} c_3^2 \right] I_a(m_K,\Delta,\mu;q^2), \label{eq:on}
\end{eqnarray}
whereas for the $\Delta^{++}$ the expression is
\begin{eqnarray}
\langle \Delta^{++} |\delta O_\mathrm{(a)}| \Delta^{++} \rangle & = & \left[ \frac58 a_1^2 + \frac54 a_1b_2 + \frac{25}{12} a_1b_3 + \frac58 b_2^2 + \frac{25}{12} b_2b_3 + \frac{125}{72} b_3^2 \right] I_a(m_\pi,0,\mu;q^2) \nonumber \\
&  & \mbox{} + \left[ \frac58 a_1^2 + \frac54 a_1b_2 + \frac{25}{12} a_1b_3 + \frac58 b_2^2 + \frac{25}{12} b_2b_3 + \frac{125}{72} b_3^2 \right] I_a(m_K,0,\mu;q^2) \nonumber \\
&  & \mbox{} + \left[ \frac12 a_1^2 + \frac12 a_1c_3 + \frac18 c_3^2 \right] I_a(m_\pi,-\Delta,\mu;q^2) + \left[ \frac12 a_1^2 + \frac12 a_1c_3 + \frac18 c_3^2 \right] I_a(m_K,-\Delta,\mu;q^2). \label{eq:odpp}
\end{eqnarray}

Equations (\ref{eq:on}) and (\ref{eq:odpp}) can be rewritten in terms of the SU(3) invariant couplings $D$, $F$, $\mathcal{C}$, and $\mathcal{H}$ introduced in heavy baryon chiral perturbation theory \cite{jm255,jm259}, which are related in the large-$N_c$ limit to the operator coefficients $a_1$, $b_2$, $b_3$, and $c_3$ of the axial vector current $A^{ia}$. The connection is given through \cite{jen96}
\begin{subequations}
\label{eq:rel1}
\begin{eqnarray}
&  & D = \frac12 a_1 + \frac16 b_3, \\
&  & F = \frac13 a_1 + \frac16 b_2 + \frac19 b_3, \\
&  & \mathcal{C} = - a_1 - \frac12 c_3, \\
&  & \mathcal{H} = -\frac32 a_1 - \frac32 b_2 - \frac52 b_3.
\end{eqnarray}
\end{subequations}

Plugging these relations into Eqs.~(\ref{eq:on}) and (\ref{eq:odpp}) yields
\begin{eqnarray}
\langle p|\delta O_\mathrm{(a)}|p\rangle & = & \frac32(D+F)^2 I_a(m_\pi,0,\mu;q^2) + (D^2 + 3 F^2) I_a(m_K,0,\mu;q^2) - \frac23 \mathcal{C}^2 I_a(m_\pi,\Delta,\mu;q^2) \nonumber \\
&  & \mbox{} + \frac16 \mathcal{C}^2 I_a(m_K,\Delta,\mu;q^2),
\end{eqnarray}
and
\begin{equation}
\langle \Delta^{++} |\delta O_\mathrm{(a)}| \Delta^{++} \rangle = \frac{5}{18} \mathcal{H}^2 I_a(m_\pi,0,\mu;q^2) + \frac{5}{18} \mathcal{H}^2 I_a(m_K,0,\mu;q^2) + \frac12 \mathcal{C}^2 I_a(m_\pi,-\Delta,\mu;q^2) + \frac12 \mathcal{C}^2 I_a(m_K,-\Delta,\mu;q^2),
\end{equation}
for $p$ and $\Delta^{++}$, respectively. 

\subsection{Diagram \ref{fig:oneloop}(b)}

The loop diagram depicted in Fig.~\ref{fig:oneloop}(b) leads to the correction \cite{rfm14b}
\begin{eqnarray}
\delta O_{\textrm{(b)}}^c & = & \frac12 \left[A^{ja},\left[A^{jb},T^c\right]\right] Q_{(1)}^{ab} - \frac12 \left\{ A^{ja}, \left[T^c,\left[\mathcal{M},A^{jb}\right] \right] \right\} Q_{(2)}^{ab} \nonumber \\
& & \mbox{} + \frac16 \left(\left[A^{ja}, \left[\left[\mathcal{M}, \left[ \mathcal{M},A^{jb}\right]\right],T^c\right] \right] - \frac12 \left[\left[\mathcal{M},A^{ja}\right], \left[\left[\mathcal{M},A^{jb}\right],T^c \right]\right]\right) Q_{(3)}^{ab} + \ldots , \label{eq:loop1b}
\end{eqnarray}
where $\mathcal{M}$ is the baryon mass operator. The contribution to the baryon Dirac form factor is simply
\begin{equation}
\delta O_{\textrm{(b)}} = \delta O_{\textrm{(b)}}^Q.
\end{equation}

The $N_c$-scaling of the expansion contained in Eq.~(\ref{eq:loop1b}) has already been discussed in detail for the baryon vector current in Ref.~\cite{rfm14b}. Because the operators $T^3$ and $T^8$ are order $\mathcal{O}(N_c^0)$ and $\mathcal{O}(N_c)$, respectively, the conclusions remain unchanged, so $\delta O_{\textrm{(b)}}$ is order $\mathcal{O}(N_c^0)$ and is of the \textit{same order} as $\delta O_{\textrm{(a)}}$.

On the other hand, $Q_{(n)}^{ab}$ in Eq.~(\ref{eq:loop1b}) is a symmetric tensor that contains the loop integral; it decomposes into flavor singlet, flavor $\mathbf{8}$, and flavor $\mathbf{27}$ representations as \cite{jen96}
\begin{eqnarray}
Q_{(n)}^{ab} = I_{b,\mathbf{1}}^{(n)} \delta^{ab} + I_{b,\mathbf{8}}^{(n)} d^{ab8} + I_{b,\mathbf{27}}^{(n)} \left[ \delta^{a8} \delta^{b8} - \frac18 \delta^{ab} - \frac35 d^{ab8} d^{888}\right], \label{eq:pisym}
\end{eqnarray}
where 
\begin{subequations}
\begin{eqnarray}
I_{b,\mathbf{1}}^{(n)} & = & \frac18 \left[3I_b^{(n)}(m_\pi,0,\mu) + 4I_b^{(n)}(m_K,0,\mu) + I_b^{(n)}(m_\eta,0,\mu) \right], \label{eq:F1} \\
I_{b,\mathbf{8}}^{(n)} & = & \frac{2\sqrt 3}{5} \left[ \frac32 I_b^{(n)}(m_\pi,0,\mu) - I_b^{(n)}(m_K,0,\mu) - \frac12 I_b^{(n)}(m_\eta,0,\mu) \right], \label{eq:F8} \\
I_{b,\mathbf{27}}^{(n)} & = & \frac13 I_b^{(n)}(m_\pi,0,\mu) - \frac43 I_b^{(n)}(m_K,0,\mu) + I_b^{(n)}(m_\eta,0,\mu). \label{eq:F27}
\end{eqnarray}
\end{subequations}

The function $I_b^{(n)}(m,0,\mu)$ corresponds to the limit $\delta \to 0$ of the general function $I_b^{(n)}(m,\delta,\mu)$ introduced in Ref.~\cite{fmhjm}; it is defined as
\begin{equation}
I_b^{(n)}(m,\delta,\mu) \equiv \frac{\partial^n I_b(m,\delta,\mu)}{\partial \delta^n}. \label{eq:fn}
\end{equation}
The full expression for $I_b(m,\delta,\mu)$ can be found in Ref.~\cite{rfm14b}. Its first three derivatives used here read
\begin{equation}
16\pi^2F_\pi^2 I_b^{(1)}(m,\delta,\mu) = (m^2-2\delta^2)\left[ \lambda_\epsilon + 1 - \ln \frac{m^2}{\mu^2} \right] - 2\delta^2 - 2\delta \sqrt{\delta^2-m^2} \ln \left[ \frac{\delta-\sqrt{\delta^2-m^2}}{\delta+\sqrt{\delta^2-m^2}} \right], \label{eq:ibp}
\end{equation}
\begin{equation}
4\pi^2F_\pi^2 I_b^{(2)}(m,\delta,\mu) = -\delta \left[ \lambda_\epsilon + 1 - \ln \frac{m^2}{\mu^2} \right] + \frac{m^2-2\delta^2}{2\sqrt{\delta^2-m^2}} \ln \left[ \frac{\delta-\sqrt{\delta^2-m^2}}{\delta+\sqrt{\delta^2-m2}}\right] \label{eq:ibpp}
\end{equation}
and
\begin{equation}
4\pi^2F_\pi^2 I_b^{(3)}(m,\delta,\mu) = -\lambda_\epsilon -\frac{\delta^2}{m^2-\delta^2} + \ln \frac{m^2}{\mu^2} + \frac{\delta}{2} \frac{3m^2-2\delta^2}{(\delta^2-m^2)^{3/2}} \ln \left[ \frac{\delta-\sqrt{\delta^2-m^2}}{\delta+\sqrt{\delta^2-m^2}}\right]. \label{eq:ibppp}
\end{equation}

With all the above ingredients, the final expression for $\delta O_{\textrm{(b)}}^c$ can be organized as
\begin{eqnarray}
\delta O_{\textrm{(b)}}^c & = & \sum_{n=1}^7 \left(c_{n}^{\mathbf{1}} S_n^c I_{b,\mathbf{1}}^{(1)} + d_n^{\mathbf{1}} S_n^c I_{b,\mathbf{1}}^{(2)} + e_n^{\mathbf{1}} S_n^c I_{b,\mathbf{1}}^{(3)} \right) + \sum_{n=1}^{13} \left(c_{n}^{\mathbf{8}} O_n^c I_{b,\mathbf{8}}^{(1)} + d_n^{\mathbf{8}} O_n^c I_{b,\mathbf{8}}^{(2)} + e_n^{\mathbf{8}} O_n^c I_{b,\mathbf{8}}^{(3)} \right) \nonumber \\
&  & \mbox{} + \sum_{n=1}^9 \left(c_{n}^{\mathbf{27}} T_n^c I_{b,\mathbf{27}}^{(1)} + d_n^{\mathbf{27}} T_n^c I_{b,\mathbf{27}}^{(2)} + e_n^{\mathbf{27}} T_n^c I_{b,\mathbf{27}}^{(3)} \right) + \ldots, \label{eq:obc}
\end{eqnarray}
where the coefficients $c_{n}^{\mathbf{r}}$, $d_{n}^{\mathbf{r}}$ and $e_{n}^{\mathbf{r}}$ and given in Eqs.~(C5)--(C13) of Ref.~\cite{rfm14b}. Notice that singlet and octet pieces must be subtracted off the $\mathbf{27}$ piece in Eq.~(\ref{eq:obc}) to have a truly $\mathbf{27}$ contribution. The operator basis for $\mathbf{27}$ flavor representations can be found in Eq.~(46) of this reference. It is nevertheless listed here for the sake of completeness; it reads,
\begin{eqnarray}
\begin{array}{ll}
T_{1}^c = f^{c8e} f^{8eg} T^g, &
T_{2}^c = f^{c8e} f^{8eg} \{J^r,G^{rg}\}, \\
T_{3}^c = f^{c8e} f^{8eg} \{J^2,T^g\}, &
T_{4}^c = \epsilon^{ijk} f^{c8e} \{G^{ke},\{J^i,G^{j8}\}\}, \\
T_{5}^c = f^{c8e} f^{8eg} \{J^2,\{J^r,G^{rg}\}\}, &
T_{6}^c = f^{c8e} f^{8eg} \{J^2,\{J^2,T^g\}\}, \\
T_{7}^c = \epsilon^{ijk} f^{c8e} \{J^2,\{G^{ke},\{J^i,G^{j8}\}\}\}, &
T_{8}^c = f^{c8e} f^{8eg} \{J^2,\{J^2,\{J^2,T^g\}\}\}, \\
T_{9}^c = \epsilon^{ijk} f^{c8e} \{J^2,\{J^2,\{G^{ke},\{J^i,G^{j8}\}\}\}\}. & \label{eq:27op}
\end{array}
\end{eqnarray}
It is important to remark that $\langle T_m^3\rangle=\langle T_m^8\rangle=0$.

By working out the two study cases of the previous section, the corresponding expressions for $p$ and $\Delta^{++}$ turn out to be
\begin{eqnarray}
\langle p|\delta O_\mathrm{(b)}|p\rangle & = & \left[ \frac{25}{24} a_1^2 + \frac{5}{12} a_1b_2 + \frac{25}{36} a_1b_3 + \frac{1}{24} b_2^2 + \frac{5}{36} b_2b_3 + \frac{25}{216} b_3^2 \right] I_b^{(1)}(m_\pi,0,\mu) \nonumber \\
&  & \mbox{} + \left[ -\frac23 a_1^2 - \frac23 a_1c_3 - \frac16 c_3^2 \right] \left[ I_b^{(1)}(m_\pi,0,\mu) + \Delta I_b^{(2)}(m_\pi,0,\mu) + \frac{\Delta^2}{2} I_b^{(3)}(m_\pi,0,\mu) \right] \nonumber \\
&  & \mbox{} + \left[ \frac{7}{12} a_1^2 + \frac13 a_1b_2 + \frac{7}{18} a_1b_3 + \frac{1}{12} b_2^2 + \frac19 b_2b_3 + \frac{7}{108} b_3^2 \right] I_b^{(1)}(m_K,0,\mu) \nonumber \\
&  & \mbox{} + \left[ \frac16 a_1^2 + \frac16 a_1c_3 + \frac{1}{24} c_3^2 \right] \left[ I_b^{(1)}(m_K,0,\mu) + \Delta I_b^{(2)}(m_K,0,\mu) + \frac{\Delta^2}{2} I_b^{(3)}(m_K,0,\mu) \right] + \ldots, \label{eq:ibpr}
\end{eqnarray}
and
\begin{eqnarray}
\langle \Delta^{++} |\delta O_\mathrm{(b)}| \Delta^{++} \rangle & = & \left[ \frac58 a_1^2 + \frac54 a_1 b_2 + \frac{25}{12} a_1b_3 + \frac58 b_2^2 + \frac{25}{12} b_2b_3 + \frac{125}{72} b_3^2 \right] I_b^{(1)}(m_\pi,0,\mu) \nonumber \\
&  & \mbox{} + \left[ \frac12 a_1^2 + \frac12 a_1c_3 + \frac18 c_3^2 \right] \left[ I_b^{(1)}(m_\pi,0,\mu) - \Delta I_b^{(2)}(m_\pi,0,\mu) + \frac{\Delta^2}{2} I_b^{(3)}(m_\pi,0,\mu) \right] \nonumber \\
&  & \mbox{} + \left[ \frac58 a_1^2 + \frac54 a_1b_2 + \frac{25}{12} a_1b_3 + \frac58 b_2^2 + \frac{25}{12} b_2b_3 + \frac{125}{72} b_3^2 \right] I_b^{(1)}(m_K,0,\mu) \nonumber \\
&  & \mbox{} + \left[ \frac12 a_1^2 + \frac12 a_1c_3 + \frac18 c_3^2 \right] \left[ I_b^{(1)}(m_K,0,\mu) - \Delta I_b^{(2)}(m_K,0,\mu) + \frac{\Delta^2}{2} I_b^{(3)}(m_K,0,\mu) \right] + \ldots \label{eq:ibde}
\end{eqnarray}
where the ellipses represent higher-order derivatives of the function $I_b(m,\delta,\mu)$. The structures of Eqs.~(\ref{eq:ibpr}) and (\ref{eq:ibde}) allow further simplifications, namely,
\begin{eqnarray}
\langle p|\delta O_\mathrm{(b)}|p\rangle & = & \left[ \frac{25}{24} a_1^2 + \frac{5}{12} a_1b_2 + \frac{25}{36} a_1b_3 + \frac{1}{24} b_2^2 + \frac{5}{36} b_2b_3 + \frac{25}{216} b_3^2 \right] I_b^{(1)}(m_\pi,0,\mu) \nonumber \\
&  & \mbox{} + \left[ \frac{7}{12} a_1^2 + \frac13 a_1b_2 + \frac{7}{18} a_1b_3 + \frac{1}{12} b_2^2 + \frac19 b_2b_3 + \frac{7}{108} b_3^2 \right] I_b^{(1)}(m_K,0,\mu) \nonumber \\
&  & \mbox{} + \left[ -\frac23 a_1^2 - \frac23 a_1c_3 - \frac16 c_3^2 \right] I_b^{(1)}(m_\pi,\Delta,\mu) + \left[ \frac16 a_1^2 + \frac16 a_1c_3 + \frac{1}{24} c_3^2 \right] I_b^{(1)}(m_K,\Delta,\mu), \label{eq:ibp2}
\end{eqnarray}
and
\begin{eqnarray}
\langle \Delta^{++} |\delta O_\mathrm{(b)}| \Delta^{++} \rangle & = & \left[ \frac58 a_1^2 + \frac54 a_1 b_2 + \frac{25}{12} a_1b_3 + \frac58 b_2^2 + \frac{25}{12} b_2b_3 + \frac{125}{72} b_3^2 \right] I_b^{(1)}(m_\pi,0,\mu) \nonumber \\
&  & \mbox{} + \left[ \frac58 a_1^2 + \frac54 a_1b_2 + \frac{25}{12} a_1b_3 + \frac58 b_2^2 + \frac{25}{12} b_2b_3 + \frac{125}{72} b_3^2 \right] I_b^{(1)}(m_K,0,\mu) \nonumber \\
&  & \mbox{} + \left[ \frac12 a_1^2 + \frac12 a_1c_3 + \frac18 c_3^2 \right] I_b^{(1)}(m_\pi,-\Delta,\mu) + \left[ \frac12 a_1^2 + \frac12 a_1c_3 + \frac18 c_3^2 \right] I_b^{(1)}(m_K,-\Delta,\mu), \label{eq:ibd2}
\end{eqnarray}
where the Maclaurin series expansion of the function $I_b^{(1)}(m,\delta,\mu)$ has been properly identified and replaced in the expressions above.

There are two key aspects worth noticing here. First, the $\eta$ meson contributions in the loop corrections (\ref{eq:ibp2}) and (\ref{eq:ibd2}) vanish, as it must be. Second, the group structure of loop diagrams from Fig.~\ref{fig:oneloop}(b) are the same as the ones from Fig.~\ref{fig:oneloop}(a), as it can be easily checked by inspecting Eqs.~(\ref{eq:ibp2})--(\ref{eq:ibd2}) and (\ref{eq:on})--(\ref{eq:odpp}). This fact will be exploited in the next section.

\begingroup
\squeezetable
\begin{table*}
\caption{\label{t:chcoeff}. Chiral coefficients for Dirac form factors.}
\begin{ruledtabular}
\begin{tabular}{lccccccc}
$B$ & $\alpha_{B}^{(\pi)}$ & $\alpha_{B}^{(K)}$ & $\beta_{B}^{(\pi)}$ & $\beta_{B}^{(K)}$ & $\gamma_B^{(\pi)}$ & $\gamma_B^{(K)}$ & $\psi_B$ \\[1mm]
\hline
$n$ & $-\frac32(D+F)^2$ & $\frac32(D-F)^2$ & $\frac23 \mathcal{C}^2$ & $\frac13 \mathcal{C}^2$ & $-\frac12$ & $\frac12$ & $-(\mathcal{C}^2 - 6DF)$ \\
$p$ & $\frac32(D+F)^2$ & $D^2+3F^2$ & $-\frac23 \mathcal{C}^2$ & $\frac16 \mathcal{C}^2$ & $\frac12$ & $1$ & $\frac12(\mathcal{C}^2 - 5D^2 - 6DF - 9 F^2)$ \\
$\Sigma^-$ & $-(D^2+3F^2)$ & $-\frac32(D-F)^2$ & $-\frac16 \mathcal{C}^2$ & $-\frac13 \mathcal{C}^2$ & $-1$ & $-\frac12$ & $\frac12(\mathcal{C}^2 + 5D^2 - 6DF + 9 F^2)$ \\
$\Sigma^0$ & $0$ & $3DF$ & $0$ & $-\frac12 \mathcal{C}^2$ & $0$ & $0$ & $\frac12(\mathcal{C}^2 - 6DF)$ \\
$\Sigma^+$ & $D^2+3F^2$ & $\frac32(D+F)^2$ & $\frac16 \mathcal{C}^2$ & $-\frac23 \mathcal{C}^2$ & $1$ & $\frac12$ & $\frac12(\mathcal{C}^2 - 5D^2 - 6DF - 9 F^2)$\\
$\Xi^-$ & $-\frac32(D-F)^2$ & $-(D^2+3F^2)$ & $-\frac13 \mathcal{C}^2$ & $-\frac16 \mathcal{C}^2$ & $-\frac12$ & $-1$ & $\frac12(\mathcal{C}^2 + 5D^2 - 6DF + 9 F^2)$\\
$\Xi^0$ & $\frac32(D-F)^2$ & $-\frac32(D+F)^2$ & $\frac13 \mathcal{C}^2$ & $\frac23 \mathcal{C}^2$ & $\frac12$ & $-\frac12$ & $-(\mathcal{C}^2 - 6DF)$\\
$\Lambda$ & $0$ & $-3DF$ & $0$ & $\frac12 \mathcal{C}^2$ & $0$ & $0$ & $-\frac12(\mathcal{C}^2 - 6DF)$ \\
$\Lambda\Sigma^0$ & $2\sqrt{3} DF$ & $\sqrt{3} DF$ & $-\frac{1}{\sqrt{3}} \mathcal{C}^2$ & $-\frac{1}{2\sqrt{3}} \mathcal{C}^2$ & $0$ & $0$ & $\frac{\sqrt{3}}{2}(\mathcal{C}^2 - 6DF)$ \\
$\Delta^{++}$ & $\frac{5}{18} \mathcal{H}^2$ & $\frac{5}{18} \mathcal{H}^2$ & $\frac12 \mathcal{C}^2$ & $\frac12 \mathcal{C}^2$ & $\frac32$ & $\frac32$ & $-\frac12(\mathcal{C}^2+\frac59\mathcal{H}^2)$ \\
$\Delta^+$ & $\frac{5}{54} \mathcal{H}^2$ & $\frac{5}{27} \mathcal{H}^2$ & $\frac16 \mathcal{C}^2$ & $\frac13 \mathcal{C}^2$ & $\frac12$ & $1$ & $-\frac12(\mathcal{C}^2+\frac59\mathcal{H}^2)$ \\
$\Delta^0$ & $-\frac{5}{54} \mathcal{H}^2$ & $\frac{5}{54} \mathcal{H}^2$ & $-\frac16 \mathcal{C}^2$ & $\frac16 \mathcal{C}^2$ & $-\frac12$ & $\frac12$ & $0$ \\
$\Delta^-$ & $ -\frac{5}{18} \mathcal{H}^2$ & $0$ & $-\frac12 \mathcal{C}^2$ & $0$ & $-\frac32$ & $0$ & $-\frac12(\mathcal{C}^2+\frac59\mathcal{H}^2)$ \\
${\Sigma^*}^+$ & $\frac{5}{27} \mathcal{H}^2$ & $\frac{5}{54} \mathcal{H}^2$ & $\frac13 \mathcal{C}^2$ & $\frac16 \mathcal{C}^2$ & $1$ & $\frac12$ & $-\frac12(\mathcal{C}^2+\frac59\mathcal{H}^2)$ \\
${\Sigma^*}^0$ & $0$ & $0$ & $0$ & $0$ & $0$ & $0$ & $0$ \\
${\Sigma^*}^-$ & $-\frac{5}{27} \mathcal{H}^2$ & $-\frac{5}{54} \mathcal{H}^2$ & $-\frac13 \mathcal{C}^2$ & $-\frac16 \mathcal{C}^2$ & $-1$ & $-\frac12$ & $-\frac12(\mathcal{C}^2+\frac59\mathcal{H}^2)$ \\
${\Xi^*}^0$ & $\frac{5}{54} \mathcal{H}^2$ & $-\frac{5}{54} \mathcal{H}^2$ & $\frac16 \mathcal{C}^2$ & $-\frac16 \mathcal{C}^2$ & $\frac12$ & $-\frac12$ & $0$ \\
${\Xi^*}^-$ & $-\frac{5}{54} \mathcal{H}^2$ & $-\frac{5}{27} \mathcal{H}^2$ & $-\frac16 \mathcal{C}^2$ & $-\frac13 \mathcal{C}^2$ & $-\frac12$ & $-1$ & $-\frac12(\mathcal{C}^2+\frac59\mathcal{H}^2)$ \\
$\Omega^-$ & $0$ & $ -\frac{5}{18} \mathcal{H}^2$ & $0$ & $-\frac12 \mathcal{C}^2$ & $0$ & $-\frac32$ & $-\frac12(\mathcal{C}^2+\frac59\mathcal{H}^2)$ 
\end{tabular}
\end{ruledtabular}
\end{table*}
\endgroup

\subsection{Diagram \ref{fig:oneloop}(c)}

The loop graph correction to the baryon Dirac form factor arising from Fig.~\ref{fig:oneloop}(c) is written as
\begin{equation}
\delta O_{\textrm{(c)}} = -if^{abe} T^e R^{ab}, \label{eq:loop1c}
\end{equation}
where $R^{ab}$ is an antisymmetric tensor which decomposes as
\begin{equation}
R^{ab} = B_0 if^{abQ} + B_1 if^{ab\overline{Q}},
\end{equation}
i.e., the flavor $\mathbf{8}$ contribution is the only one present in $R^{ab}$, whereas the flavor $\mathbf{10} + \mathbf{\overline{10}}$ contribution is absent. Besides, the integral over the loop, $I_c(m,\mu;q^2)$, is contained in the coefficients $B_0$ and $B_1$ as
\begin{subequations}
\label{eq:ait}
\begin{eqnarray}
B_0 & = & \frac13 [ I_c(m_\pi,\mu;q^2)+2I_c(m_K,\mu;q^2) ], \\
B_1 & = & \frac13 [ I_c(m_\pi,\mu;q^2)-I_c(m_K,\mu;q^2) ],
\end{eqnarray}
\end{subequations}
where
\begin{equation}
16\pi^2 F_\pi^2 I_c(m,\mu;q^2) = \left[ m^2+\frac16 q^2 \right]\left[ -\lambda_\epsilon - 1 + \ln \frac{m^2}{\mu^2} \right] - \frac43 m^2 - \frac{5}{18} q^2 - \frac{[q^2(4m^2+q^2)]^{3/2}}{6(q^2)^2} \ln \left[ \frac{-q^2+\sqrt{q^2(4m^2+q^2)}}{q^2+\sqrt{q^2(4m^2+q^2)}} \right]. \label{eq:ic}
\end{equation}

A further calculation yields
\begin{equation}
\delta O_{\textrm{(c)}} = N_f ( B_0 T^Q + B_1 T^{\overline{Q}} ), \label{eq:loop1c2}
\end{equation}
where $N_f$ is the number of light quark flavors. Thus, $\delta O_{\textrm{(c)}}$ is also order $\mathcal{O}(N_c^0)$, as in the previous two cases.

\subsection{Diagram \ref{fig:oneloop}(d)}

Finally, the structure for the Feynman diagram of Fig.~\ref{fig:oneloop}(d) is \cite{rfm14b}
\begin{equation}
\delta O_{\textrm{(d)}}^c = - \frac12 \left[T^a,\left[T^b,T^c\right]\right] S^{ab}, \label{eq:loop1d}
\end{equation}
where $S^{ab}$ is a symmetric tensor given by
\begin{eqnarray}
S^{ab} = I_{d,\mathbf{1}} \delta^{ab} + I_{d,\mathbf{8}} d^{ab8} + I_{d,\mathbf{27}} \left[ \delta^{a8} \delta^{b8} - \frac18 \delta^{ab} - \frac35 d^{ab8} d^{888}\right], \label{eq:gamsym}
\end{eqnarray}
where
\begin{subequations}
\begin{eqnarray}
I_{d,\mathbf{1}} & = & \frac18 \left[3I_d(m_\pi,\mu) + 4I_d(m_K,\mu) + I_d(m_\eta,\mu) \right], \label{eq:G1} \\
I_{d,\mathbf{8}} & = & \frac{2\sqrt 3}{5} \left[ \frac32 I_d(m_\pi,\mu) - I_d(m_K,\mu) - \frac12 I_d(m_\eta,\mu) \right], \label{eq:G8} \\
I_{d,\mathbf{27}} & = & \frac13 I_d(m_\pi,\mu) - \frac43 I_d(m_K,\mu) + I_d(m_\eta,\mu). \label{eq:G27}
\end{eqnarray}
\label{eq:loopggs}
\end{subequations}

The loop integral $I_d(m,\mu)$ is now written as
\begin{equation}
I_d(m,\mu) = \frac{m^2}{16\pi^2F_\pi^2}\left[ -\lambda_\epsilon - 1 + \ln{\frac{m^2}{\mu^2}} \right]. \label{eq:id}
\end{equation}

The flavor structures contained in Eq.~(\ref{eq:loop1d}) are easily evaluated. They read \cite{rfm14b} \\

(1) Flavor singlet contribution
\begin{equation}
[T^a,[T^a,T^c]] = N_f T^c. \label{eq:sind}
\end{equation}

(2) Flavor $\mathbf{8}$ contribution
\begin{equation}
d^{ab8} [T^a,[T^b,T^c]] = \frac{N_f}{2} d^{c8e} T^e. \label{eq:octd}
\end{equation}

(3) Flavor $\mathbf{27}$ contribution
\begin{equation}
[T^8,[T^8,T^c]] = f^{c8e} f^{8eg} T^g. \label{eq:27d}
\end{equation}

Therefore, the contribution from Fig.~\ref{fig:oneloop}(d) to the Dirac form factor is
\begin{equation}
\delta O_{\textrm{(d)}} = \delta O_{\textrm{(d)}}^Q,
\end{equation}
which indeed is also $\mathcal{O}(N_c^0)$. Contributions $\delta O_{\textrm{(c)}}$ and $\delta O_{\textrm{(d)}}$ can be combined into a single expression and this fact will also be exploited in the next section.

\section{\label{sec:totalc}Total correction to the Dirac form factor}

In the conventional chiral momentum counting scheme, aside from one-loop corrections to the Dirac form factor, tree diagrams involving higher-order vertices also contribute \cite{jen92,kubis2001,geng09}. They come along with low-energy constants (LECs) so the number of unknowns in the low-energy expansion increases. Up to order $\mathcal{O}(p^3)$, there is not any unknown LEC that comes into play: at order $\mathcal{O}(p^2)$ there is one LEC associated, at least for decuplet baryons, to the SU(3) symmetric description of the anomalous part on the magnetic moment, while at order $\mathcal{O}(p^3)$, there are two LECs that describe a SU(3)-symmetric part of the electric quadrupole moment and charge radius \cite{arn03,geng09}. At one-loop order, however, there are two more LECs that play the role of local counterterms for the divergent parts of the integrals; they will be referred to as $\zeta_1$ and $\zeta_2$ hereafter.

Gathering together all partial results (tree plus one-loop terms), the Dirac form factor $F_{1,B}(q^2)$ for baryon $B$ can be written in the compact form as
\begin{equation}
F_{1,B}(q^2) = Q_B \left[1 - \frac{q^2}{\Lambda_\chi^2} \zeta_1 \right] - \frac{q^2}{\Lambda_\chi^2} \psi_B \zeta_2 + \sum_{\phi=\pi,K} \left[ \alpha_B^{(\phi)} \mathcal{R}_1(m_\phi,0,\mu;q^2) + \beta_B^{(\phi)} \mathcal{R}_1(m_\phi,\Delta,\mu;q^2) + \gamma_B^{(\phi)} \mathcal{R}_2(m_\phi,\mu;q^2)\right] \label{eq:f1bo}
\end{equation}
for octet baryons and
\begin{equation}
F_{1,B}(q^2) = Q_B \left[1 - \frac{q^2}{\Lambda_\chi^2} (\zeta_1 + \psi_B \zeta_2) \right] + \sum_{\phi=\pi,K} \left[ \alpha_B^{(\phi)} \mathcal{R}_1(m_\phi,0,\mu;q^2) + \beta_B^{(\phi)} \mathcal{R}_1(m_\phi,-\Delta,\mu;q^2) + \gamma_B^{(\phi)} \mathcal{R}_2(m_\phi,\mu;q^2)\right] \label{eq:f1bd}
\end{equation}
for decuplet baryons, where the coefficients $\alpha_B^{(\phi)}$, $\beta_B^{(\phi)}$, and $\psi_B$ for baryon $B$, with $\phi=\pi,K$, are listed in Table \ref{t:chcoeff}, written in terms of the SU(3) invariant couplings, and $\Lambda_\chi=4\pi F_\pi$. Furthermore, $\mathcal{R}_1(m,\delta,\mu;q^2) \equiv I_a(m,\delta,\mu;q^2) + I_b^{(1)}(m,\delta,\mu)$ and $\mathcal{R}_2(m,\mu;q^2) \equiv I_c(m,\mu;q^2) - I_d(m,\mu)$. The explicit forms of these functions are
\begin{eqnarray}
16\pi^2 F_\pi^2\mathcal{R}_1(m,\delta,\mu;q^2) & = & \frac{1}{18}q^2 \left[ - \frac{14}{3} + \ln \frac{m^2}{\mu^2} \right] - \frac{16}{9}m^2 + 4\delta^2 - 2\delta \sqrt{\delta^2-m^2} \ln \left[ \frac{\delta-\sqrt{\delta^2-m^2}}{\delta+\sqrt{\delta^2-m^2}} \right] \nonumber \\
&  & \mbox{} - \frac{16m^2+q^2-36\delta^2}{18q^2} \sqrt{q^2(4m^2+q^2)} \ln \left[ \frac{-q^2+\sqrt{q^2(4m^2+q^2)}}{q^2+\sqrt{q^2(4m^2+q^2)}} \right] \nonumber \\
&  & \mbox{} - \int_0^1 dx \frac{2\delta}{\sqrt{\delta^2-q^2(1-x)x-m^2}} \left[ m^2-\delta^2 + \frac43 q^2(1-x)x \right] \ln \left[ \frac{\delta-\sqrt{\delta^2-q^2(1-x)x-m^2}}{\delta+\sqrt{\delta^2-q^2(1-x)x-m^2}} \right], \nonumber \\
\end{eqnarray}
and
\begin{equation}
16\pi^2 F_\pi^2\mathcal{R}_2(m,\mu;q^2) = \frac16 q^2 \left[ - \frac83 + \ln \frac{m^2}{\mu^2} \right] - \frac43 m^2 - \frac{[q^2(4m^2+q^2)]^{3/2}}{6(q^2)^2} \ln \left[ \frac{-q^2+\sqrt{q^2(4m^2+q^2)}}{q^2+\sqrt{q^2(4m^2+q^2)}} \right].
\end{equation}

For consistency, in the limit $q^2\to 0$,
\begin{equation}
\lim_{q^2\to 0}\mathcal{R}_1(m,\delta,\mu;q^2) = 0,
\end{equation}
and
\begin{equation}
\lim_{q^2\to 0} \mathcal{R}_2(m,\mu;q^2) = 0,
\end{equation}
which follow from
\begin{equation}
\lim_{q^2\to 0} I_a(m,\delta,\mu;q^2) = -I_b^{(1)}(m,\delta,\mu),
\end{equation}
and
\begin{equation}
\lim_{q^2\to 0} I_c(m,\mu;q^2) = I_d(m,\mu),
\end{equation}
respectively. Thus, as expected, the one-loop correction vanishes in the limit $q^2\to 0$.

An important number of relations among Dirac form factor can be tested using Eqs.~(\ref{eq:f1bo}) and (\ref{eq:f1bd}), inspired by the original expressions introduced for baryon magnetic moments. Coleman and Glashow \cite{cg61} derived some useful relations valid in the SU(3) limit. Thus, for the Dirac form factor, the Coleman and Glashow-like relations read
\begin{eqnarray}
\begin{array}{lcl}
F_{1,{\Sigma^+}}(q^2) - F_{1,p}(q^2) = \sum_i k_{1i}, & \qquad \qquad & F_{1,{\Sigma^-}}(q^2) + F_{1,n}(q^2) + F_{1,p}(q^2) = \sum_i k_{2i}, \\[4mm]
2F_{1,\Lambda}(q^2) - F_{1,n}(q^2) = \sum_i k_{3i}, & \qquad \qquad & F_{1,{\Xi^-}}(q^2) - F_{1,{\Sigma^-}}(q^2) = \sum_i k_{4i}, \\[4mm]
F_{1,{\Xi^0}}(q^2) - F_{1,n}(q^2) = \sum_i k_{5i}, & \qquad \qquad & 2F_{1,{\Lambda\Sigma^0}}(q^2) + \sqrt{3}F_{1,n}(q^2) = \sum_i k_{6i},
\end{array}
\end{eqnarray}
where the various $k_{ji}$ are functions that depend quadratically on the SU(3) invariant couplings and the differences $\mathcal{R}_l(m_K)-\mathcal{R}_l(m_\pi)$. They are due to flavor representations other than octet [Figs.~\ref{fig:oneloop}(a) and (c)] and singlet [Figs.~\ref{fig:oneloop}(b) and (d)].

The pioneering work by Caldi and Pagels \cite{caldi} on baryon magnetic moments using chiral perturbation theory found some sum rules that are valid up to one-loop corrections of order $\mathcal{O}(m_q^{1/2})$. The equivalent expressions for the Dirac form factor read
\begin{equation}
F_{1,\Sigma^+}(q^2) + 2F_{1,\Lambda}(q^2) + F_{1,\Sigma^-}(q^2) = 0, \label{eq:cp1}
\end{equation}
\begin{equation}
F_{1,\Xi^0}(q^2) + F_{1,\Xi^-}(q^2) + F_{1,n}(q^2) - 2F_{1,\Lambda}(q^2) + F_{1,p}(q^2) = 0, \label{eq:cp2}
\end{equation}
and
\begin{equation}
F_{1,\Lambda}(q^2) - \sqrt{3} F_{1,\Lambda\Sigma^0}(q^2) - F_{1,\Xi^0}(q^2) - F_{1,n}(q^2) = 0. \label{eq:cp3}
\end{equation}

Jenkins \textit{et.\ al} \cite{jen92}, in the framework of heavy baryon chiral perturbation theory, found an expression among baryon magnetic moments valid including all terms of orders $\mathcal{O}(m_q^{1/2})$, $\mathcal{O}(m_q\ln m_q)$, and $\mathcal{O}(m_q)$. The equivalent expression involving Dirac form factor reads
\begin{equation}
6 F_{1,\Lambda}(q^2) + F_{1,\Sigma^-}(q^2) - 4 \sqrt{3} F_{1,\Lambda\Sigma^0}(q^2) - 4 F_{1,n}(q^2) + F_{1,\Sigma^+}(q^2) - 4 F_{1,\Xi^0}(q^2) = 0.
\end{equation}

Lebed and Martin \cite{lb} found other combinations among baryon magnetic moments sensitive to $I=2$ and $I=3$. Those expressions are also satisfied using Eqs.~(\ref{eq:f1bo}) and (\ref{eq:f1bd}), i.e., for $I=2$
\begin{equation}
F_{1,\Sigma^+}(q^2) - 2F_{1,\Sigma^0}(q^2) + F_{1,\Sigma^-}(q^2) = 0, \label{eq:is1}
\end{equation}
\begin{equation}
F_{1,\Delta^{++}}(q^2) - F_{1,\Delta^+}(q^2) - F_{1,\Delta^0}(q^2) + F_{1,\Delta^-}(q^2) = 0,
\end{equation}
\begin{equation}
F_{1,{\Sigma^*}^+}(q^2) - 2 F_{1,{\Sigma^*}^0}(q^2) + F_{1,{\Sigma^*}^-}(q^2) = 0,
\end{equation}
whereas for $I=3$
\begin{equation}
F_{1,\Delta^{++}}(q^2) - 3 F_{1,\Delta^+}(q^2) + 3 F_{1,\Delta^0}(q^2) - F_{1,\Delta^-}(q^2) = 0.
\end{equation}

There are additional relations to be tested. As it was mentioned in the introductory section, the combined formalism was applied to evaluate two flavor $\mathbf{27}$ combinations of baryon masses in Ref.~\cite{jen96}. One of them is the Gell-Mann--Okubo combination for baryon octet masses,
\begin{equation}
\frac34 M_\Lambda + \frac14 M_\Sigma - \frac12M_N - \frac12 M_\Xi, \label{eq:gmoC}
\end{equation}
and the other one is the decuplet equal spacing rule,
\begin{equation}
-\frac47 M_\Delta + \frac57 M_{\Sigma^*} + \frac27 M_{\Xi^*} - \frac37 M_\Omega. \label{eq:esrC}
\end{equation}

There are eight linear combinations of Dirac form factors that transform as $I=0$. They are
\begin{subequations}
\begin{eqnarray}
F_{1,N_0}(q^2) & = & \frac12 [F_{1,n}(q^2) + F_{1,p}(q^2)], \\
F_{1,\Sigma_0}(q^2) & = & \frac13 [F_{1,\Sigma^+}(q^2) + F_{1,\Sigma^0}(q^2) + F_{1,\Sigma^-}(q^2)], \\
F_{1,\Xi_0}(q^2) & = & \frac12 [F_{1,\Xi^0}(q^2) + F_{1,\Xi^-}(q^2)], \\
F_{1,\Delta_0}(q^2) & = & \frac14 [F_{1,\Delta^{++}}(q^2) + F_{1,\Delta^+}(q^2) + F_{1,\Delta^0}(q^2) + F_{1,\Delta^-}(q^2)], \\
F_{1,\Sigma_0^*}(q^2) & = & \frac13 [F_{1,{\Sigma^*}^+}(q^2) + F_{1,{\Sigma^*}^0}(q^2) + F_{1,{\Sigma^*}^-}(q^2)], \\
F_{1,\Xi_0^*}(q^2) & = & \frac12 [F_{1,{\Xi^*}^0}(q^2) + F_{1,{\Xi^*}^-}(q^2)],
\end{eqnarray}
\end{subequations}
along with $F_{1,\Lambda}(q^2)$ and $F_{1,\Omega^-}(q^2)$.

A direct substitution of the above expressions into relations (\ref{eq:gmoC}) and (\ref{eq:esrC}) yields
\begin{equation}
\frac34 F_{1,\Lambda}(q^2) + \frac14 F_{1,\Sigma_0}(q^2) - \frac12 F_{1,N_0}(q^2) - \frac12 F_{1,\Xi_0}(q^2) = 0, \label{eq:gmo}
\end{equation}
and
\begin{equation}
-\frac47 F_{1,\Delta_0}(q^2) + \frac57 F_{1,{\Sigma_0^*}}(q^2) + \frac27 F_{1,{\Xi_0^*}}(q^2) - \frac37 F_{1,\Omega^-}(q^2) = 0. \label{eq:esr}
\end{equation}

It is important to emphasize that the above relations among Dirac form factors are valid up to order $\mathcal{O}(p^2)$ in the chiral counting. Any violations of them should arise from higher-order terms.

\section{\label{sec:bcr}Baryon charge radii}

An important static property that can be derived from the Dirac form factor is the mean-square charge radius $\langle r_B^2\rangle$ of a baryon $B$; it is defined by the relation
\begin{equation}
\langle r_B^2 \rangle = -\frac{6}{Q_B} \left. \frac{d}{dq^2} G_{E0}(q^2) \right|_{q^2=0}, \label{eq:cr2def}
\end{equation}
while for neutral baryons, the normalization factor $1/Q_B$ in Eq.~(\ref{eq:cr2def}) is simply dropped.

From definitions (\ref{eq:sachs1}) and (\ref{eq:sachs1d}), $\langle r_B^2\rangle$ reads
\begin{equation}
\langle r_B^2 \rangle = -\frac{6}{Q_B} \left[ \left. \frac{d}{dq^2} F_{1,B}(q^2) \right|_{q^2=0} - \frac{1}{4M_B^2} F_2(0) \right], \label{eq:cr2osac}
\end{equation}
and
\begin{equation}
\langle r_B^2 \rangle = -\frac{6}{Q_B} \left[ \left. \frac{d}{dq^2} F_{1,B}(q^2) \right|_{q^2=0} - \frac{1}{4M_B^2} F_2(0) + \frac{1}{6M_B^2}G_{E2}(0) \right], \label{eq:cr2dsac}
\end{equation}
for octet and decuplet baryons, respectively.

Now, from Eqs.~(\ref{eq:f1bo}) and (\ref{eq:f1bd}), the expressions for $\langle r_B^2 \rangle$ follow easily. They read
\begin{equation}
\langle r_B^2 \rangle = -\frac{6}{Q_B} \left[ - \frac{1}{\Lambda_\chi^2} (Q_B \zeta_1 + \psi_B \zeta_2) + \zeta_3 + \sum_{\phi=\pi,K} \left[ \alpha_B^{(\phi)} \mathcal{R}_1^\prime(m_\phi,0,\mu) + \beta_B^{(\phi)} \mathcal{R}_1^\prime(m_\phi,\Delta,\mu) + \gamma_B^{(\phi)} \mathcal{R}_2^\prime(m_\phi,\mu)\right] \right], \label{eq:cro}
\end{equation}
for octet baryons and
\begin{equation}
\langle r_B^2 \rangle = -\frac{6}{Q_B} \left[ - \frac{1}{\Lambda_\chi^2} Q_B (\zeta_1 + \psi_B \zeta_2) + \zeta_4 + \sum_{\phi=\pi,K} \left[ \alpha_B^{(\phi)} \mathcal{R}_1^\prime(m_\phi,0,\mu) + \beta_B^{(\phi)} \mathcal{R}_1^\prime(m_\phi,-\Delta,\mu) + \gamma_B^{(\phi)} \mathcal{R}_2^\prime(m_\phi,\mu)\right] \right], \label{eq:crd}
\end{equation}
for decuplet baryons, where
\begin{equation}
\zeta_3 = \frac{3}{2M_B^2} F_2(0), \qquad \qquad \zeta_4 = \frac{3}{2M_B^2} F_2(0) - \frac{1}{M_B^2}G_{E2}(0).
\end{equation}

The primed $\mathcal{R}_l$ functions are easily obtained as
\begin{eqnarray}
\mathcal{R}_1^\prime(m,\delta,\mu) & \equiv & \left. \frac{d}{dq^2} \mathcal{R}_1(m,\delta,\mu;q^2)\right|_{q^2=0} \nonumber \\
& = & \frac{1}{288\pi^2 F_\pi^2} \times \left\{ \begin{array}{ll} 
\displaystyle \ln\frac{m^2}{\mu^2} - \frac{5\delta}{\sqrt{\delta^2-m^2}} \ln \left[ \frac{\delta-\sqrt{\delta^2-m^2}}{\delta+\sqrt{\delta^2-m^2}} \right] & |\delta| > m \\[6mm]
\displaystyle \ln\frac{m^2}{\mu^2} + \frac{10\delta}{\sqrt{m^2-\delta^2}} \left[ \frac{\pi}{2} - \tan^{-1} \left[\frac{\delta}{\sqrt{m^2-\delta^2}} \right] \right], & |\delta| < m,
\end{array}
\right.
\end{eqnarray}
and
\begin{eqnarray}
\mathcal{R}_2^\prime(m,\mu) & \equiv & \left. \frac{d}{dq^2} \mathcal{R}_2(m,\mu;q^2)\right|_{q^2=0} \nonumber \\
& = & \frac{1}{96\pi^2 F_\pi^2} \ln\frac{m^2}{\mu^2}.
\end{eqnarray}

As is the case with Dirac form factors, some useful relations among mean-square charge radii can also be found. Because no expressions for electric quadrupole moments have been computed within the combined formalism yet, those terms are excluded in the following relations. Also, explicit SU(3) symmetry breaking in the magnetic moments is removed from the original expressions introduced in Ref.~\cite{rfm14}. Thus, with proper adjustment of negative charges, the relations found among mean-square charge radii, including tree and one-loop contributions only, are the following:

For the Coleman and Glashow-like relations, one finds
\begin{eqnarray}
\begin{array}{lcl}
\langle r_{\Sigma^+}^2 \rangle - \langle r_p^2 \rangle = h_1, & \qquad \qquad & \langle r_{\Sigma^-}^2 \rangle + \langle r_n^2 \rangle + \langle r_p^2 \rangle = h_2, \\[4mm]
2\langle r_\Lambda^2 \rangle - \langle r_n^2 \rangle = h_3, & \qquad \qquad & \langle r_{\Xi^-}^2 \rangle - \langle r_{\Sigma^-}^2 \rangle = h_4, \\[4mm]
\langle r_{\Xi^0}^2 \rangle - \langle r_n^2 \rangle = h_5, & \qquad \qquad & 2\langle r_{\Lambda\Sigma^0}^2 \rangle + \sqrt{3}\langle r_n^2 \rangle = h_6, \label{eq:cg}
\end{array}
\label{eq:treeval}
\end{eqnarray}
where the $h_i$'s on the right-hand sides of Eq.~(\ref{eq:cg}) are given functions of the SU(3) invariant couplings and meson masses. Their precise forms are not needed here; suffice it to say that all the $h_i$ functions vanish in the SU(3) limit.

As for the Caldi and Pagels-like relations, one gets
\begin{equation}
\langle r^2_{\Sigma^+} \rangle + 2 \langle r^2_{\Lambda} \rangle - \langle r^2_{\Sigma^-} \rangle = 0,
\end{equation}
\begin{equation}
\langle r^2_{\Xi^0} \rangle - \langle r^2_{\Xi^-} \rangle + \langle r^2_{n} \rangle - 2 \langle r^2_{\Lambda} \rangle + \langle r^2_{p} \rangle = 0,
\end{equation}
and
\begin{equation}
\langle r^2_{\Lambda} \rangle - \sqrt{3} \langle r^2_{\Lambda\Sigma^0} \rangle - \langle r^2_{\Xi^0} \rangle - \langle r^2_{n} \rangle = 0.
\end{equation}

Also, the Jenkins \textit{et.\ al}--like relation reads
\begin{equation}
6 \langle r^2_{\Lambda} \rangle - \langle r^2_{\Sigma^-} \rangle - 4 \sqrt{3} \langle r^2_{\Lambda\Sigma^0} \rangle - 4 \langle r^2_{n} \rangle + \langle r^2_{\Sigma^+} \rangle - 4 \langle r^2_{\Xi^0} \rangle = 0.
\end{equation}

Additionally, Lebed and Buchmann \cite{buch03} found other combinations among mean-square charge radii sensitive to $I=2$ and $I=3$. Those expressions are also satisfied using Eqs.~(\ref{eq:cro}) and (\ref{eq:crd}), i.e.,
\begin{equation}
2 \langle r^2_{\Delta^{++}} \rangle - 3 \langle r^2_{\Delta^+} \rangle + 3 \langle r^2_{\Delta^0} \rangle + \langle r^2_{\Delta^-} \rangle = 0,
\end{equation}
\begin{equation}
\langle r^2_{{\Sigma^*}^+} \rangle - 2 \langle r^2_{{\Sigma^*}^0} \rangle - \langle r^2_{{\Sigma^*}^-} \rangle = 0,
\end{equation}
\begin{equation}
\langle r^2_{\Sigma^+} \rangle - 2 \langle r^2_{\Sigma^0} \rangle - \langle r^2_{\Sigma^-} \rangle = 0,
\end{equation}
for $I=2$, and
\begin{equation}
2 \langle r^2_{\Delta^{++}} \rangle - \langle r^2_{\Delta^+} \rangle - \langle r^2_{\Delta^0} \rangle - \langle r^2_{\Delta^-} \rangle = 0,
\end{equation}
for $I=3$.

In a close parallelism to the previous section, for the $I=0$ charge radius relations it is possible to define
\begin{subequations}
\begin{eqnarray}
\langle r_{N_0}^2\rangle & = & \frac12 [\langle r_{n}^2 \rangle + \langle r_{p}^2 \rangle], \\
\langle r_{\Sigma_0}^2\rangle & = & \frac13 [\langle r_{\Sigma^+}^2 \rangle + \langle r_{\Sigma^0}^2 \rangle - \langle r_{\Sigma^-}^2 \rangle], \\
\langle r_{\Xi_0}^2\rangle & = & \frac12 [\langle r_{\Xi^0}^2 \rangle - \langle r_{\Xi^-}^2 \rangle], \\
\langle r_{\Delta_0}^2\rangle & = & \frac14 [2\langle r_{\Delta^{++}}^2 \rangle + \langle r_{\Delta^+}^2 \rangle + \langle r_{\Delta^0}^2 \rangle - \langle r_{\Delta^-}^2 \rangle], \\
\langle r_{\Sigma_0^*}^2\rangle & = & \frac13 [\langle r_{{\Sigma^*}^+}^2 \rangle + \langle r_{{\Sigma^*}^0}^2 \rangle - \langle r_{{\Sigma^*}^-}^2 \rangle], \\
& = & 0, \\ 
\langle r_{\Xi_0^*}^2\rangle & = & \frac12 [\langle r_{{\Xi^*}^0}^2 \rangle - \langle r_{{\Xi^*}^-}^2 \rangle],
\end{eqnarray}
\end{subequations}

A direct substitution of the above expressions into relations (\ref{eq:gmoC}) and (\ref{eq:esrC}) yields
\begin{equation}
\frac34 \langle r_{\Lambda}^2 \rangle + \frac14 \langle r_{\Sigma_0}^2 \rangle - \frac12 \langle r_{N_0}^2 \rangle - \frac12 \langle r_{\Xi_0}^2 \rangle = f_B(\mu_i), \label{eq:gmoCR}
\end{equation}
and
\begin{equation}
-\frac47 \langle r_{\Delta_0}^2 \rangle + \frac57 \langle r_{{\Sigma_0^*}}^2 \rangle + \frac27 \langle r_{{\Xi_0^*}}^2 \rangle + \frac37 \langle r_{\Omega^-}^2 \rangle = f_T(\mu_i), \label{eq:esrCR}
\end{equation}
where $f_B$ and $f_T$ are linear functions of the magnetic moments of the octet and decuplet baryons, respectively. These results are the expected ones because the magnetic moments do not satisfy relations (\ref{eq:gmoC}) and (\ref{eq:esrC}).
Needless to say, the above relations between mean-square charge radii are valid up to order $\mathcal{O}(p^2)$ in the chiral counting.

It is now instructive to perform a comparison of the results obtained here with others presented in the literature. Unfortunately, theoretical works on the subject within heavy baryon chiral perturbation theory are rather scarce. References \cite{puglia} and \cite{arn03} provide expressions for octet and decuplet baryons, respectively, so they can be used for comparative purposes. In the first case, charge radii are computed to order $\mathcal{O}(1/\Lambda_\chi M_N)$, including decuplet states in the loops; in the second case, although the formalisms used are quenched and partially quenched chiral perturbation theory (at next-to-leading order in the chiral expansion, leading order in the heavy baryon expansion), results in chiral perturbation theory are presented too.

One-loop corrections to the charge radii of Ref.~\cite{puglia} can be written as
\begin{equation}
\langle r_B^2\rangle \sim -\sum_{\phi=\pi,K} \left[ \beta^{(\phi)} F(m_\phi,0,\mu) + {\beta^\prime}^{(\phi)} F(m_\phi,\Delta,\mu) \right], \label{eq:tibcr}
\end{equation}
where the chiral coefficients $\beta^{(\phi)}$ and ${\beta^\prime}^{(\phi)}$ are listed in the appendix and $F(m_\phi,\Delta,\mu)$ is the loop integral given in Eq.~(6) of this reference.

Equation (\ref{eq:cro}) can be adapted to Eq.~(\ref{eq:tibcr}) as
\begin{equation}
\langle r_B^2\rangle \sim -\sum_{\phi=\pi,K} \left[ \left( \frac{1}{18} \alpha_B^{(\phi)} + \frac16 \gamma_B^{(\phi)} \right) F(m_\phi,0,\mu) + \frac{1}{18} \beta_B^{(\phi)} F^\prime(m_\phi,\Delta,\mu) \right]. \label{eq:cro2}
\end{equation}
Thus, there should be a matching between the quantities $\frac{1}{18} \alpha_B^{(\phi)} + \frac16 \gamma_B^{(\phi)}$ and $ \beta^{(\phi)}$ on the one hand and $\frac{1}{18} \beta_B^{(\phi)}$ and ${\beta^\prime}^{(\phi)}$ on the other hand. In the former case the matching does not occur whereas in the latter case it does, but the structures of the loop integral are rather different beyond certain point; this is indicated by putting a prime on $F(m_X,\delta,\mu)$ in Eq.~(\ref{eq:cro2}).

A crucial test of Eq.~(\ref{eq:tibcr}) would be to check whether it fulfills the sum rules among charge radii discussed above. The answer is no.

As for charge radii of decuplet baryons, Eq.~(37) of Ref.~\cite{arn03} provides the expression to compare with. The one-loop correction can be written as
\begin{equation}
\langle r_B^2\rangle \sim -\frac13 \frac{9+5\mathcal{C}^2}{\Lambda_\pi^2} \sum_\phi A_\phi G(m_\phi,0,\mu) - \frac{25}{27} \frac{\mathcal{H}^2}{\Lambda_\chi^2} \sum_\phi A_\phi G(m_\phi,\Delta,\mu), \label{eq:at}
\end{equation}
where the function $G(m_\phi,\Delta,\mu)$ is explicitly given in that Eq.~(37). Equation (\ref{eq:crd}) could also be rewritten to contrast it with expression (\ref{eq:at}), but it turns out to be meaningless. The answer is simple. The coupling constants $\mathcal{C}$ and $\mathcal{H}$ parametrize the vertices $\phi BT$ and $\phi TT$, respectively. Therefore, terms proportional to $\mathcal{H}^2$ and $\mathcal{C}^2$ come along with loop diagrams with $\Delta=0$ and $\Delta\neq 0$, respectively; this fact is not satisfied by Eq.~(\ref{eq:at}). It also fails to fulfill the appropriate sum rules discussed above. Therefore, the comparison between Eqs.~(\ref{eq:crd}) and (\ref{eq:at}) cannot be carried on.

An additional result that allows a partial comparison is the one presented in Ref.~\cite{geng09}, where the mean-square charge radii of decuplet baryons in covariant chiral perturbation were calculated. The comparison can be performed at the level of chiral coefficients of corresponding diagrams, listed in Table VIII of this reference. Except for global factors, the coefficients agree in full.

Further comparisons with other approaches will be performed numerically in the next section.

\section{\label{sec:num}Numerical analysis}

It is now instructive to produce some numbers in order to check how Eqs.~(\ref{eq:cro}) and (\ref{eq:crd}) work. The authors of Ref.~\cite{part} quote the values of three measured mean-square charge radii,\footnote{Actually, except for $n$, the quoted values are those of the charge radii.} specifically,
\begin{subequations}
\label{eq:crexp}
\begin{eqnarray}
\langle r^2_{n} \rangle & = & -0.1161 \pm 0.0022 \\
\langle r^2_{p} \rangle & = & 0.70706 \pm 0.00066 \,\,(\mu p \,\,\, \mathrm{Lamb shift}), \qquad 0.770 \pm 0.009 \,\,\,(ep \, \, \mathrm{CODATA}) \\
\langle r^2_{\Sigma^-} \rangle & = & 0.608 \pm 0.156.
\end{eqnarray}
\end{subequations}

For definiteness, the constants involved in Eqs.~(\ref{eq:cro}) and (\ref{eq:crd}) are fixed as follows: the numerical values of the meson masses are $m_\pi = 0.13957$ and $m_K= 0.49368 \,\,\mathrm{GeV}$ \cite{part}. For baryon masses, $M_B=1.151$ and $M_T=1.382$, so $\Delta = 0.231\,\,\mathrm{GeV}$. Finally, $F_\pi = 0.093$ and $\mu=1\,\,\mathrm{GeV}$. Equations (\ref{eq:cro}) and (\ref{eq:crd}) also depend on the SU(3) invariant couplings, two free parameters $\zeta_1$ and $\zeta_2$ (counterterms), and two more parameters $\zeta_3$ and $\zeta_4$; by excluding the electric quadrupole moment, $\zeta_4$ along with $\zeta_3$ are given in terms of the baryon anomalous magnetic moments and in principle they can be evaluated \cite{rfm14}. Therefore, the invariant couplings and counterterms have to be pinned down from some other sources. The experimental data, on the other hand, are not enough to attempt a global fit to extract all the free parameters. A cautious way to approach the puzzle is to pick a model that provides values of $D$, $F$, $\mathcal{C}$, and $\mathcal{H}$ and to fit for $\zeta_1$ and $\zeta_2$. For this, at least two possible scenarios can be addressed:
\begin{enumerate}
\item[(a)] To use the nonrelativistic quark model predictions, namely,
\begin{equation}
\frac{F}{D} = \frac23, \qquad \mathcal{C} = -2D, \qquad \mathcal{H} = -3D, \label{eq:qm}
\end{equation}
and then fixing $D$ by means of $F+D$ = 1.27, which corresponds to the value of the axial coupling in neutron $\beta$ decay. This yields $D=0.76$, $F=0.51$, $\mathcal{C}=-1.52$, and $\mathcal{H}=-2.29$. Furthermore, the use of expressions (\ref{eq:qm}) reduces by one the free parameters, so the combination $-(Q_B \zeta_1 + \psi_B \zeta_2)/\Lambda_\chi^2$ can actually be written as $-Q_B\zeta_1^\prime/\Lambda_\chi^2$.
\item[(b)] To use $D$, $F$, $\mathcal{C}$, and $\mathcal{H}$ extracted from the analysis of the vector and axial vector form factors of BSD in the framework of the combined chiral and $1/N_c$ expansions in Ref.~\cite{rfm14b}, namely, $D=0.66$, $F=0.25$, $\mathcal{C}=-1.48$, and $\mathcal{H}=-2.50$.
\end{enumerate}

Armed with the invariant couplings, the operator coefficients $a_1$, $b_2$, $b_3$, and $c_3$ can easily be obtained by inverting relation (\ref{eq:rel1}); afterwards, they can be used to feed the theoretical expressions to evaluate the baryon magnetic moments with the formalism of Ref.~\cite{rfm14}. The analyses of Refs.~\cite{rfm14} and \cite{rfm14b} have been performed under the same footing, so all the predictions should be consistent. A word of caution is in order at this point. The electric quadruple moments are not accounted for in the following numerical analyses. These quantities come along with terms that are formally of order $\mathcal{O}(p^3)$ and in principle should be suppressed compared to the ones retained.

Without further ado, fitting to the experimental data yields for scenario (a)
\begin{equation}
\zeta_1^\prime = 2.34 \pm 0.75,
\end{equation}
where in order to get a meaningful fit a theoretical error of $0.156\,\,\mathrm{fm}^2$ has been added in quadrature; even so $\chi^2=1.6/\mathrm{dof}$, so the badness of the fit is reflected directly into the large error of $\zeta_1^\prime$.

For scenario (b), the fit yields
\begin{equation}
\zeta_1 = 1.65 \pm 0.05, \qquad \zeta_2 = -1.04 \pm 0.01,
\end{equation}
with $\chi^2= 0.66/\mathrm{dof}$ with no theoretical error added. The $\zeta_i$ parameters so obtained are in accord with expectations. They are at most order $\mathcal{O}(N_c)$ as they originate from one-loop contributions only.

\begingroup
\squeezetable
\begin{table*}
\caption{\label{t:crpre}. Predicted mean-square charge radii for baryons, depending on whether the SU(3) invariant couplings are obtained (a) via the nonrelativistic quark model relations or (b) with SU(3) breaking effects evaluated within the chiral and $1/N_c$ expansions \cite{rfm14,rfm14b}.}
\begin{ruledtabular}
\begin{tabular}{lrrrrrrrrr}
& \multicolumn{4}{c}{Scenario (a)} & \multicolumn{5}{c}{Scenario (b)} \\
$B$ & Total & $\zeta^\prime_1$ & $\zeta_3$ & One loop & Total & $\zeta_1$ & $\zeta_2$ & $\zeta_3$ & One loop \\[1mm]
\hline
$n$               & $-0.385$ & $ 0.000$ & $-0.085$ & $-0.300$ & $-0.116$ & $ 0.000$ & $ 0.211$ & $-0.085$ & $-0.242$ \\
$p$               & $ 0.812$ & $ 0.528$ & $ 0.079$ & $ 0.205$ & $ 0.770$ & $ 0.284$ & $ 0.135$ & $ 0.079$ & $ 0.271$ \\
$\Sigma^-$        & $ 0.524$ & $ 0.528$ & $ 0.007$ & $-0.011$ & $ 0.735$ & $ 0.284$ & $ 0.346$ & $ 0.007$ & $ 0.097$ \\
$\Sigma^0$        & $ 0.092$ & $ 0.000$ & $ 0.029$ & $ 0.063$ & $-0.026$ & $ 0.000$ & $-0.106$ & $ 0.029$ & $ 0.051$ \\
$\Sigma^+$        & $ 0.708$ & $ 0.528$ & $ 0.064$ & $ 0.116$ & $ 0.682$ & $ 0.284$ & $ 0.135$ & $ 0.064$ & $ 0.199$ \\
$\Xi^-$           & $ 0.416$ & $ 0.528$ & $-0.015$ & $-0.097$ & $ 0.649$ & $ 0.284$ & $ 0.346$ & $-0.015$ & $ 0.034$ \\
$\Xi^0$           & $-0.183$ & $ 0.000$ & $-0.055$ & $-0.128$ & $ 0.059$ & $ 0.000$ & $ 0.211$ & $-0.056$ & $-0.097$ \\
$\Lambda$         & $-0.090$ & $ 0.000$ & $-0.027$ & $-0.063$ & $ 0.029$ & $ 0.000$ & $ 0.106$ & $-0.026$ & $-0.051$ \\
$\Lambda\Sigma^0$ & $ 0.281$ & $ 0.000$ & $ 0.070$ & $ 0.211$ & $ 0.050$ & $ 0.000$ & $-0.183$ & $ 0.066$ & $ 0.166$ \\
$\Delta^{++}$     & $ 0.657$ & $ 0.528$ & $ 0.091$ & $ 0.038$ & $ 1.048$ & $ 0.284$ & $ 0.498$ & $ 0.091$ & $ 0.176$ \\
$\Delta^+$        & $ 0.726$ & $ 0.528$ & $ 0.144$ & $ 0.055$ & $ 1.101$ & $ 0.284$ & $ 0.498$ & $ 0.130$ & $ 0.189$ \\
$\Delta^0$        & $ 0.138$ & $ 0.000$ & $ 0.105$ & $ 0.033$ & $ 0.105$ & $ 0.000$ & $ 0.000$ & $ 0.078$ & $ 0.027$ \\
$\Delta^-$        & $ 0.450$ & $ 0.528$ & $-0.066$ & $-0.012$ & $ 0.891$ & $ 0.284$ & $ 0.498$ & $-0.026$ & $ 0.136$ \\
${\Sigma^*}^+$    & $ 0.440$ & $ 0.528$ & $-0.109$ & $ 0.021$ & $ 0.939$ & $ 0.284$ & $ 0.498$ & $-0.004$ & $ 0.162$ \\
${\Sigma^*}^0$    & $-0.009$ & $ 0.000$ & $-0.009$ & $ 0.000$ & $-0.031$ & $ 0.000$ & $ 0.000$ & $-0.031$ & $ 0.000$ \\
${\Sigma^*}^-$    & $ 0.543$ & $ 0.528$ & $-0.007$ & $ 0.021$ & $ 0.895$ & $ 0.284$ & $ 0.498$ & $-0.049$ & $ 0.162$ \\
${\Xi^*}^0$       & $-0.240$ & $ 0.000$ & $-0.206$ & $-0.033$ & $-0.098$ & $ 0.000$ & $ 0.000$ & $-0.071$ & $-0.027$ \\
${\Xi^*}^-$       & $ 0.585$ & $ 0.528$ & $ 0.002$ & $ 0.055$ & $ 0.981$ & $ 0.284$ & $ 0.498$ & $ 0.010$ & $ 0.189$ \\
$\Omega^-$        & $ 0.661$ & $ 0.528$ & $ 0.045$ & $ 0.088$ & $ 1.042$ & $ 0.284$ & $ 0.498$ & $ 0.045$ & $ 0.216$
\end{tabular}
\end{ruledtabular}
\end{table*}
\endgroup

\begin{figure}[t]
\scalebox{0.85}{\includegraphics{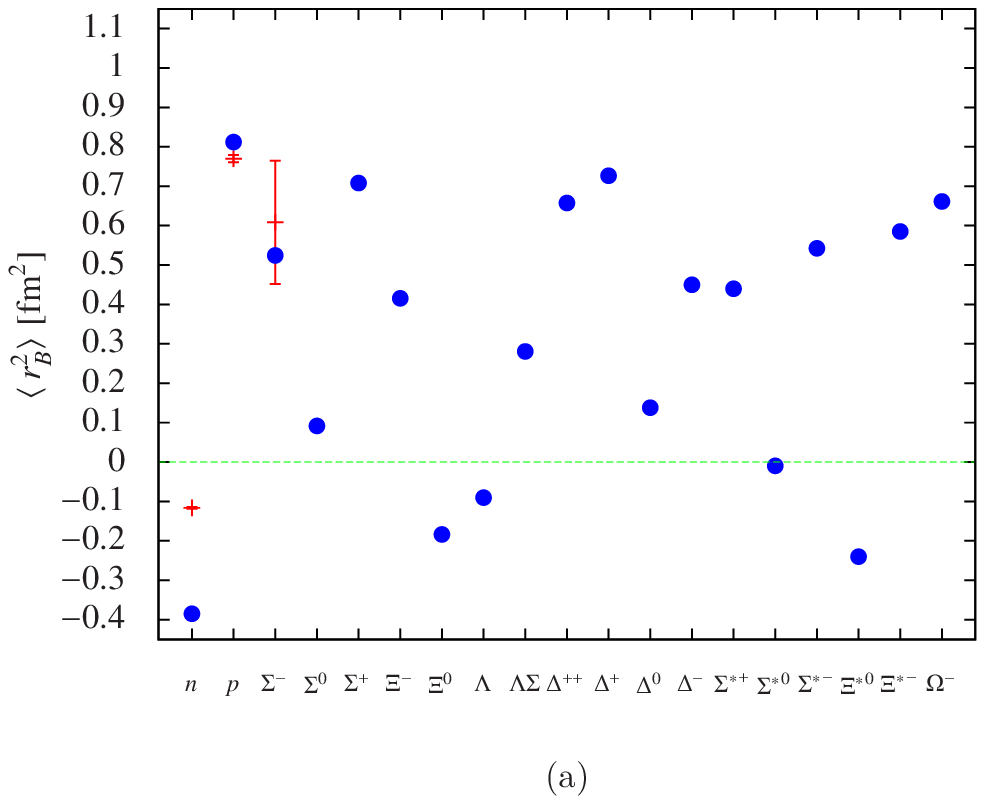}}
\scalebox{0.85}{\includegraphics{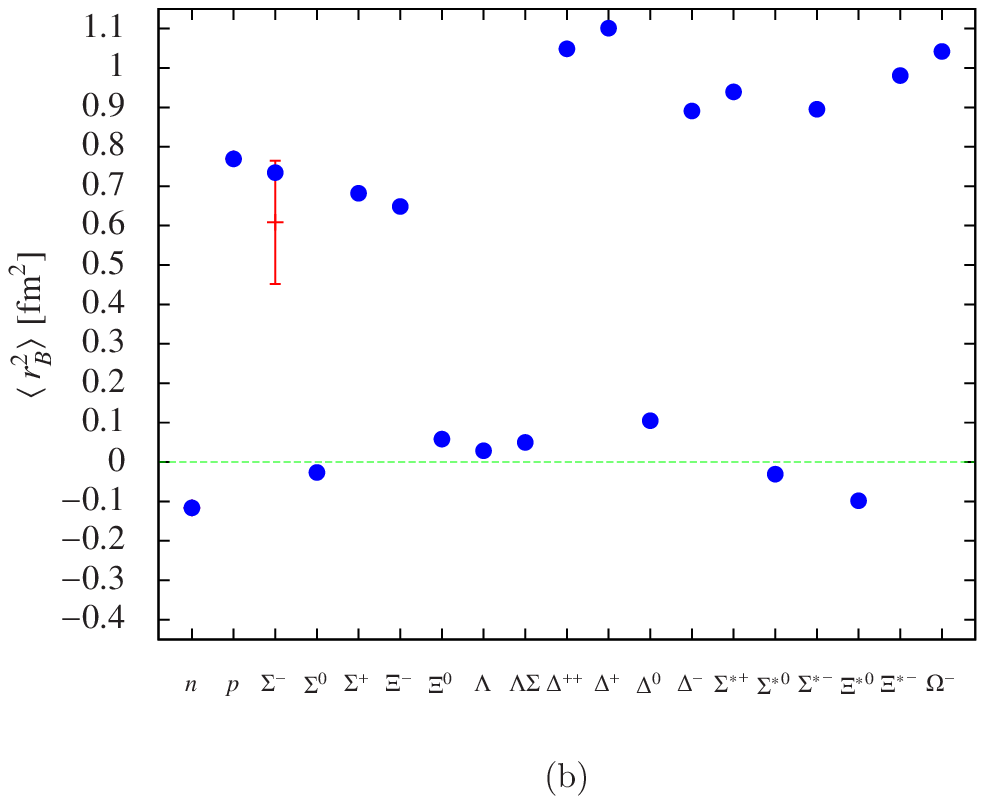}}
\caption{\label{fig:crpre}Mean-square charge radii for octet and decuplet baryons for the SU(3) invariant couplings determined in two possible scenarios: (a) The nonrelativistic quark model and (b) The chiral and $1/N_c$ expansions. The filled circles represent predictions. The three observed mean-square charge radii are also plotted.}
\end{figure}

With the above partial results, all possible $\langle r^2_B \rangle$ are listed in Table \ref{t:crpre} and plotted in Fig.~\ref{fig:crpre} for scenarios (a) and (b) for completeness. In such a table, the different contributions that make up $\langle r^2_B \rangle$ are listed as well.

Some interesting features emanate from these objects. As mentioned above, the $\langle r^2_B \rangle$ corresponding to scenario (a) were evaluated under the assumption that the SU(3) invariant couplings could be obtained from their SU(6) predictions. This turned out to be an awkward assumption. After all, the charge radii have been computed at one-loop order within the combined formalism, which already includes an implicit breaking of flavor SU(3) symmetry through the meson masses in the loops. It turns out that the $\langle r^2_B \rangle$ spread out all over the region as it can be better appreciated in Fig.~\ref{fig:crpre}(a).

In scenario (b), the SU(3) invariant couplings used are those obtained on an equal footing from the analyses of baryon magnetic moments \cite{rfm14} and axial and vector couplings \cite{rfm14b}. As a result, some patterns emerge: First, all $\langle r^2_B \rangle$ corresponding to neutral baryons lie around zero, which is consistent with $G_{E0}(0)=0$. In other words, such particles can emit virtual photons and possess a charge distribution, which would explain these nonvanishing charge radii. Second, the $\langle r^2_B \rangle$ of both octet and decuplet charged baryons reveal an apparent hierarchy and therefore they are concentrated in two well-defined narrow bands, the one for decuplet baryons lying higher than its counterpart for octet baryons. This is a reasonable result because one would expect that decuplet baryons be more extended in space than octet baryons.

At a glance, the entries of Table \ref{t:crpre} indicate that the different contributions that make up $\langle r^2_B \rangle$ are indeed as expected from the large-$N_c$ and chiral countings. Loop contributions are as important as tree-level ones. In particular, for the neutron, there is a large cancellation between the $\zeta_2$ and the one-loop terms so that $\langle r^2_n \rangle$ gets most of its value from $\zeta_3$ (here $\zeta_3$ corresponds, up to normalization factor, to the so-called Foldy term defined for the nucleon). This term is also responsible, in the present approach, for a nonzero $\langle r^2_{{\Sigma^*}^0} \rangle$, which otherwise would vanish.

The total values of $\langle r^2_B \rangle$ for scenario (b) of Table \ref{t:crpre} can be compared with some other results presented in the literature. Unfortunately, most papers are focused to either octet or decuplet baryons so it is quite hard to assess the success of the different calculations. There is a single paper \cite{dahiya} that provides values of charge radii for both octet and decuplet baryons in the context of the constituent chiral quark model. By comparing the different values, the agreement observed is very good on general grounds, except for the fact that $\langle r^2_{\Omega^-}\rangle$ is roughly one third of the value reported here. This is worrisome, especially because a recent lattice simulation \cite{ale10} found $\langle r^2_{\Omega^-} \rangle= 0.328\mathrm{-}0.355\,\,\mathrm{fm}^2$ for a pion mass or around $0.35\,\,\mathrm{GeV}$, which is far from the prediction reported here. The numerical analysis was performed using the measured mean-square charge radii of $n$, $p$, and $\Sigma^-$. In scenario (b), the first two values are well reproduced whereas the latter, within the experimental error, is fairly well reproduced. If this $\langle r^2_{\Omega^-} \rangle$ from lattice were also included as data in the fit, the $\chi^2$ would be too high to represent an output with a coherent physical content.

Systematic lower values of $\langle r^2_B \rangle$ are also observed in recent theoretical works in the context of a manifest covariant effective field theory \cite{led} and a covariant Bethe-Salpeter approach \cite{san}.

Within the combined approach, corrections from tree, one-loop and counterterms are under reasonable control and are in accord with expectations from the quiral and $1/N_c$ countings. The predictions, however, are not the ultimate in the sense that more theoretical work is required to incorporate $\mathcal{O}(p^3)$ contributions. This represents a formidable task and will be attempted in the near future. Adding this contribution, perhaps, could remedy the discrepancy observed in $\langle r^2_{\Omega^-} \rangle$. Of course, new or improved data will be also welcome in the near future.

\section{\label{sec:rem}Concluding remarks}

In this paper the Dirac form factors and consequently the mean-square charge radii of baryons have been computed at one-loop level in a combined formalism in chiral and $1/N_c$ corrections to order $\mathcal{O}(p^2)$ in the usual chiral counting. The $1/N_c$ chiral effective Lagrangian for the lowest-lying baryons was constructed in Ref.~\cite{jen96} and describes the interactions of the spin-$\frac12$ baryon octet and the spin-$\frac32$ baryon decuplet with the pion nonet. This formalism has been applied successfully to other baryon static properties, particularly magnetic moments \cite{rfm09,rfm14} and axial and vector couplings \cite{rfm14b}. For the Dirac form factor, much of the work already advanced in the latter reference could be borrowed and adapted to the present analysis.

The Dirac form factors were thus constructed at one-loop level plus tree-level contributions that play a crucial role in the analysis. By working to order $\mathcal{O}(p^2)$ in the chiral counting and practically at all orders in the $1/N_c$ expansion, it was possible to write the final expressions in terms of the SU(3) invariant couplings $D$, $F$, $\mathcal{C}$, and $\mathcal{H}$ introduced in heavy baryon chiral perturbation theory \cite{jm255,jm259}. These couplings, along with the anomalous magnetic moments, are the necessary inputs to find the charge radii unambiguously. Two possible sets of couplings are used here. One set is dictated by SU(6) symmetry and the other set, which has a better physical content, is obtained on the same footing as in the present paper \cite{rfm14,rfm14b}. The former case, referred to as scenario (a) here, yielded a quite poor fit. As a result, the charge radii were also poorly determined. In contrast, the latter case, referred to as scenario (b), yielded a good fit with very good predictions for charge radii.

In passing, it is worth mentioning that several relations among Dirac form factors and charge radii found in the literature are also nicely fulfilled with the results presented here. This provided an extra cross-check.

The decuplet-octet mass difference has been taken into account, but neglected the SU(3) splittings of the octet and decuplet baryons. The combined formalism to order $\mathcal{O}(p^2)$ represents a first step towards a more refined calculation where these splittings can be included to the next order, $\mathcal{O}(p^3)$. This inclusion, however, requires a non-negligible effort.

\acknowledgments

This work has been partially supported by Consejo Nacional de Ciencia y Tecnolog{\'\i}a and Fondo de Apoyo a la Investigaci\'on (Universidad Aut\'onoma de San Luis Potos{\'\i}), Mexico.

\end{document}